\documentclass[sigconf]{acmart}

\AtBeginDocument{%
  }

\copyrightyear{2023} 
\acmYear{2023} 
\setcopyright{rightsretained} 
\acmConference[FPGA '23]{Proceedings of the 2023 ACM/SIGDA International Symposium on Field Programmable Gate Arrays}{February 12--14, 2023}{Monterey, CA, USA}
\acmBooktitle{Proceedings of the 2023 ACM/SIGDA International Symposium on Field Programmable Gate Arrays (FPGA '23), February 12--14, 2023, Monterey, CA, USA}
\acmDOI{10.1145/3543622.3573182}
\acmISBN{978-1-4503-9417-8/23/02}

\usepackage[normalem]{ulem}

\usepackage{amsmath,amsfonts}
\usepackage{algorithm}
 \usepackage{algpseudocode}
 \usepackage{algorithmicx}
\usepackage{algcompatible}
\usepackage{graphicx}
\usepackage{textcomp}
\usepackage{xcolor}
\usepackage{xspace}
\usepackage{comment}

\usepackage{wrapfig}

\algrenewcommand\algorithmicrequire{\textbf{Input:}}
\algrenewcommand\algorithmicensure{\textbf{Output:}}

\definecolor{peekcolor}{rgb}{0.760, 0.890, 1}
\definecolor{readcolor}{gray}{0.888}

\usepackage{newfloat}
\DeclareFloatingEnvironment[
    fileext=los,
    listname={List of Code},
    name=Code,
    placement=tbhp,
]{mycode}

\newcommand{\authorsqwe}{Linghao Song, 
Licheng Guo,
Suhail Basalama,
Yuze Chi,
Robert F. Lucas,
and
Jason Cong}

\ccsdesc[500]{Hardware~Hardware accelerators}
\ccsdesc[500]{Computer systems organization~Reconfigurable computing}

\keywords{High Bandwidth Memory, Conjugate Gradient, Accelerator.}

\newcommand{\calli}{\textsc{Callipepla}\xspace}

\begin{document}

\title{\calli: Stream Centric Instruction Set and 
Mixed Precision for Accelerating Conjugate Gradient Solver}


\author[Linghao Song, 
Licheng Guo,
Suhail Basalama,
Yuze Chi,
Robert F. Lucas,
and
Jason Cong]{Linghao Song, 
Licheng Guo,
Suhail Basalama,
Yuze Chi,
Robert F. Lucas$^\dagger$,
and
Jason Cong
}

\affiliation{%
  \institution{University of California,  Los Angeles \hspace{0.2cm} $^\dagger$Livermore Software Technology,
an ANSYS Company}
  \country{}
}
\email{{linghaosong,lcguo,basalama,
chiyuze,cong}@cs.ucla.edu,bob.lucas@ansys.com}


\begin{abstract}

The continued growth in the processing power of FPGAs coupled with high bandwidth memories (HBM), makes systems like the Xilinx U280 credible platforms for linear solvers which often dominate the run time of scientific and engineering applications. In this paper, we present \calli, an accelerator for a preconditioned conjugate gradient linear solver (CG).  FPGA acceleration of CG faces three challenges: (1) how to support an arbitrary problem and terminate acceleration processing on the fly, (2) how to coordinate long-vector data flow among processing modules, and (3) how to save off-chip memory bandwidth and maintain double (FP64) precision accuracy. To tackle the three challenges, we present (1) a stream-centric instruction set for efficient streaming processing and control, (2) vector streaming 
reuse (VSR) and decentralized vector flow scheduling to coordinate vector data flow among modules and further reduce off-chip memory accesses with a double memory channel design, and (3) a mixed precision scheme to save bandwidth yet still achieve effective double precision quality solutions. 
To the best of our knowledge,
this is the first work to introduce the
concept of VSR for data reusing between 
on-chip modules to reduce unnecessary off-chip accesses for FPGA accelerators.
We prototype the accelerator on a Xilinx U280 HBM FPGA. Our evaluation shows that compared to the Xilinx HPC product, the
XcgSolver, \calli achieves a speedup of $3.94\times$, $3.36\times$ higher throughput, and $2.94\times$ better energy efficiency.
Compared to an NVIDIA A100 GPU which has $4\times$ the memory bandwidth of \calli, we still achieve 77\% of its throughput with $3.34\times$ higher energy efficiency.
The code is available at \url{https://github.com/UCLA-VAST/Callipepla}.
\end{abstract}


\maketitle

\vspace{-6pt}
\section{Introduction}
\label{sec:intro}
The need to solve large systems of linear 
equations is common in scientific and 
engineering fields
including mathematics,
physics, chemistry, and other natural sciences
subjects \cite{ferziger2002computational,griewank2008evaluating,antoulas2005approximation}, and often dominates 
their runtime.  As these linear systems 
are often sparse, it is not practical to 
invert them, and the requirements for 
storage and time grow superlinearly 
if one tries to factor them. Therefore, 
for well conditioned problems, practitioners 
are drawn to iterative algorithms whose 
storage requirements are minimal, yet still 
converge to a useful solution in a reasonable 
period of time.

The Conjugate Gradients method~\cite{hestenes1952methods}
is a well known iterative solver. 
Coupled with even the simplest Jacobi 
preconditioner (JPCG)~\cite{saad2003iterative}, it is 
very effective for solving linear 
systems that are symmetric and positive definite.
The acceleration of (preconditioned)
conjugate gradient on general-purpose
platforms CPUs and GPUs suffer from 
low computational efficiency~\cite{nakano1997parallel,helfenstein2012parallel}. 
The computational efficiency is
even worse for distributed and clustered
computers~\cite{dongarra2015hpcg}.

The recent High Bandwidth Memory (HBM)
equipped FPGAs~\cite{u280}
enable us to customize accelerator
architecture and optimize data
flows with high memory bandwidth
for the conjugate gradient solver.
In this work, we present \calli, an accelerator prototyped on Xilinx U280
HBM FPGA for Jacobi preconditioned conjugate gradient linear solver.
We resolve three challenges in 
FPGA CG acceleration with our
innovative solutions.

\textbf{Challenge 1: } The support 
of an arbitrary problem and accelerator
termination on the fly.
The FPGA
synthesis/place/route
flow takes hours even days to complete, which prevents the frequent invocation
of conjugate gradient solvers on different problems in data centers.
Thus, we need to build an accelerator 
to support an arbitrary problem.
However,
the JPCG has six dimensions of freedom
as illustrated in Algorithm~\ref{alg:cg}, which
makes 
it challenging for the JPCG accelerator to support
an arbitrary problem. 
Furthermore, it is difficult
to terminate the accelerator 
on the fly for a preset threshold (Line 6 of Algorithm~\ref{alg:cg}),
because we do not know
when to terminate until
we run the algorithm/accelerator.
To resolve this challenge, 
we design a stream centric instruction set
to control the processing modules and 
data flows in the accelerators.
We encode vector and matrix
size and data flow directions in the
instruction. 
A global controller is responsible
to issue instructions.
Therefore, we are able
so support an arbitrary problem for JPCG
and
terminate the accelerator.

\textbf{Challenge 2: } The coordination
of long-vector data flow among processing modules. JPCG involves
the processing of multiple
long vectors whose size is larger
than on-chip memory.
The latency cost is high if we always
write a produced vector to
off-chip memory and read a vector from
off-chip memory when a module consumes
the vector.
There are opportunities to save off-chip memory read and write by 
reusing vectors among processing modules.
For this challenge, we introduce 
the concept of vector streaming 
reuse and
analyze the dependencies in JPCG to decide
when and which vectors we may reused
by processing modules via on-chip
streaming and store/load the other vectors that can not be reused to/from
off-chip memory. Based on the
vector reusing/loading/storing analysis, 
we form a decentralized vector scheduling
to dissolve the global control to vector control and computation modules
to coordinate vector
flows among modules.

\textbf{Challenge 3: } Saving off-chip memory bandwidth and maintaining double (FP64) precision convergence.
The sparse matrix with FP64 precision
dominates memory footprint and thus
memory bandwidth in the sparse matrix vector multiply (SpMV)
of JPCG.
Lower precision (less bits per data element) provides
higher parallelism.
However, lower precision leads to 
larger iteration count for convergence
or even failure to converge.
To resolve this issue,
we present a mixed FP32/FP64
precision for SpMV in JPCG
to save memory bandwidth and
achieve effective convergence as default
FP64 precision.

Our evaluation Xilinx U280 HBM FPGAs shows that compared to the Xilinx HPC product XcgSolver, \calli achieves a speedup of $3.94\times$, $3.36\times$ higher throughput, and $2.94\times$ better energy efficiency.
Compared to an NVIDIA A100 GPU which has $4\times$ the memory bandwidth of \calli, we still achieve 
77\% of its throughput with $3.34\times$ higher energy efficiency.

\vspace{-12pt}
\section{CG Solver Acceleration Challenges \& \calli Solutions}
\label{sec:back}
\vspace{-3pt}
\subsection{Conjugate Gradient Solver}
\begin{minipage}{0.95\columnwidth}
\vspace{-12pt}
\begin{algorithm}[H]
\footnotesize
\caption{Jacobi preconditioner conjugate gradient solver for solving a linear system $\mathbf{A} \vec{\mathbf{x}}= \vec{\mathbf{b}}$.} 
\label{alg:cg} 
\begin{algorithmic}[0]
\REQUIRE ~~
(1) matrix $\mathbf{A}$, 
(2) Jacobi preconditioner $\mathbf{M}$,
(3) reference vector $\vec{\mathbf{b}}$, 
(4) initial solution vector $\vec{\mathbf{x}}_0$, 
(5) convergence threshold $\tau$, and
(6) maximum iteration number $N_{\text{max}}$.

\ENSURE ~~
A solution vector $\vec{\mathbf{x}}$.
\\
\end{algorithmic}
\vspace{-9pt}
\begin{algorithmic}[1]

\STATE $\vec{\mathbf{r}} \gets \vec{\mathbf{b}} - \mathbf{A} \vec{\mathbf{x}}_0$

\STATE $\vec{\mathbf{z}} \gets \mathbf{M}^{-1} \vec{\mathbf{r}}$

\STATE $\vec{\mathbf{p}} \gets \vec{\mathbf{z}}$

\STATE $\text{rz} \gets \vec{\mathbf{r}}^{\top}\cdot\vec{\mathbf{z}}$

\STATE $\text{rr} \gets \vec{\mathbf{r}}^{\top}\cdot\vec{\mathbf{r}}$

\FOR{($0\leq i < N_{\text{max}}$ and $\text{rr} > \tau$)} 

\STATE $\vec{\mathbf{ap}} \gets \mathbf{A} \vec{\mathbf{p}}$

\STATE $\alpha \gets \text{rz} / (\vec{\mathbf{p}}^{\top}\cdot\vec{\mathbf{ap}})$

\STATE $\vec{\mathbf{x}} \gets \vec{\mathbf{x}} +
\alpha\vec{\mathbf{p}}$

\STATE $\vec{\mathbf{r}} \gets \vec{\mathbf{r}} -
\alpha\vec{\mathbf{ap}}$

\STATE $\vec{\mathbf{z}} \gets \mathbf{M}^{-1} \vec{\mathbf{r}}$

\STATE $\text{rz\_new} \gets \vec{\mathbf{r}}^{\top}\cdot\vec{\mathbf{z}}$

\STATE $\vec{\mathbf{p}} \gets \vec{\mathbf{z}} +
(\text{rz\_new}/\text{rz})\vec{\mathbf{p}}$

\STATE $\text{rz} \gets \text{rz\_new}$

\STATE $\text{rr} \gets \vec{\mathbf{r}}^{\top}\cdot\vec{\mathbf{r}}$

\ENDFOR

\end{algorithmic}
\end{algorithm}
\vspace{-9pt}
\end{minipage}

For non-trivial problems, it is not practical nor hardware
efficient to directly inverse
the matrix $\mathbf{A}$ to solve
the linear system because
$\mathbf{A}$ can be very large in
real-world applications.
The conjugate gradient~\cite{hestenes1952methods}
iteratively refines errors and reaches
the solution. 
The Jacobi preconditioner~\cite{saad2003iterative}
approximates the matrix with its diagonal, which is trivial to invert.
Using even this simple Jacobi preconditioner helps reduce 
the iteration 
number and accelerate the conjugate gradient method.

Algorithm~\ref{alg:cg} illustrates the
Jacobi preconditioned conjugate gradient algorithm (JPCG) 
for solving the linear system $\mathbf{A} \vec{\mathbf{x}}= \vec{\mathbf{b}}$. 
JPCG takes as input the matrix $\mathbf{A}$, 
the Jacobi preconditioner $\mathbf{M}$, i.e.,
the diagonal of $\mathbf{A}$,
a reference vector $\vec{\mathbf{b}}$, 
an initial solution vector $\vec{\mathbf{x}}_0$, 
convergence threshold $\tau$, and
a maximum iteration count $N_{\text{max}}$.
In the algorithm,
$\vec{\mathbf{r}}$ represents the error
of current solution vector $\vec{\mathbf{x}}$,
the cooperation of vector $\vec{\mathbf{z}}$,
and vector $\vec{\mathbf{p}}$ helps the
solution vector $\vec{\mathbf{x}}$ refine
to the correct values at each iteration,
and vector $\vec{\mathbf{ap}}$ is the product
of matrix $\mathbf{A}$ and vector $\vec{\mathbf{p}}$. 
Line 1 to 3 compute the initial 
values of 
the vectors $\vec{\mathbf{r}}$,
$\vec{\mathbf{z}}$, and $\vec{\mathbf{p}}$
given the initial solution vector 
$\vec{\mathbf{x}}_0$.
Lines 4, 5 initialize two scalars,
rz and rr. In the main loop body,
the JPCG updates the vectors 
and scalars.
Note that for the computation of 
$\vec{\mathbf{z}} \gets \mathbf{M}^{-1}  \vec{\mathbf{r}}$, because $\mathbf{M}$ is
a diagonal matrix, the invert and multiply 
operation becomes an element-wise divide.
In summary, the JPCG involves the coordination of multiple kernels. The sparse matrix vector multiplication
\texttt{SpMV}, dot product, and generalized vector addition \texttt{axpy} are the core
computations. 
Note, any practical implementation of the JPCG, whether on a GPU or and FPGA, will frequently invoke memory load operations to fetch the matrix and the vectors, as well as stores on the vectors. This is because the size of the linear system dwarfs the on-chip memory capacity.

We accelerate the JPCG in this work
because:

\noindent$\bullet$ The JPCG is an important
solver used in the
industry. For example, 
the JPCG is a solver in Ansys LS-DYNA~\cite{lsdyna}, a finite element program
for engineering simulation. In Xilinx
Vitis HPC Libraries~\cite{hpcvitis}, 
the JPCG is the only linear system solver.

\noindent$\bullet$ The JPCG is hardware efficient.
There are more powerful preconditioners one could consider, which generally reduce the number of iterations required to solver the linear system.
For example, incomplete Cholesky factorization
(ICCG)~\cite{kershaw1978incomplete}
employs the lower triangular matrix from an incomplete Cholesky factorization
as the preconditioner matrix $\mathbf{M} = \mathbf{L}$.
But solving $\vec{\mathbf{z}} \gets \mathbf{L}^{-1} \vec{\mathbf{r}}$
incurs massive dependency issues,
thus, difficult to process in parallel in hardware.
We will address that in future work.

\vspace{-15pt}
\subsection{Prior CG Acceleration \& Related Works}
\noindent
\textbf{CG acceleration.}
\cite{maslennikow2005fpga}~implemented the basic CG on FPGAs while the problem dimension supported is less than 1,024, 
thus, 
not practical for real-world applications. \cite{lopes2008high}~implemented a floating-point basic CG
for dense matrices.
The supported matrix dimension is less than 100, and 
it 
needed to generate a new
hardware accelerator
for every new problem instance. 
\cite{roldao2009more}~explored reducing FP mantissa bits to reduce FP computation latency and resource. 
\cite{rampalli2018efficient}~implemented CG for Laplacian systems. However, \cite{roldao2009more,rampalli2018efficient} stored vectors all on chip without off-chip memory optimization and was limited
to small size problems. Even the minimum throughput achieved by \calli is 1.35$\times$ higher than the maximum throughput achieved by \cite{rampalli2018efficient}. Besides, 
the reduced bit method of \cite{roldao2009more} 
led to considerable iteration gaps
compared with default FP64, but our mixed-precision scheme makes the gap negligible. The GMRES leverages low precision in error correction computation and the inner loop~\cite{higham2022mixed} to reduce processing time. It is unexplored how to co-optimize mixed precision with memory accessing on the modern HBM FPGAs.
All the prior works~\cite{maslennikow2005fpga,lopes2008high,roldao2009more,rampalli2018efficient} did not optimize off-chip memory accesses, did not leverage preconditioners to accelerate the convergence, were not able to terminate accelerator on the fly, and were not able to perform large scale CG (where the matrix dimension can be
a few hundred thousands to several millions). XcgSolver~\cite{hpcvitis,hpcvitis_github}
by Xilinx is a state-of-the-art FPGA CG solver which can run real-world large-scale CG. So we use XcgSolver as the baseline.

\noindent
\textbf{Other related works.} 
GraphR~\cite{song2018graphr} and GraphLily~\cite{hu2021graphlily}
are accelerators based on non-traditional (other than DDR) memories
for graph processing. Sextans~\cite{song2022sextans} and Serpens~\cite{song2022serpens}
are SpMM/SpMV accelerators on HBM memories, while Fowers et al.~\cite{fowers2014high} designed an SpMV accelerator on DDR memory. 
Song et al.~\cite{song2020low} explored
mixed precision for conjugate gradient solvers in ReRAM.
TAPA~\cite{chi2021extending}
provides a framework for 
task-parallel FPGA programing,
and AutoBridge~\cite{guo2021autobridge} optimized the floor planning for high level synthesis and thus boosted the frequency of generated accelerators~\cite{guo2022tapa}.
Cheng et al.~\cite{cheng2020combining,cheng2021dass} explored the combination of
static and dynamic scheduling in high-level synthesis (HLS). 
Coarse grain reconfigurable architecture (CGRA) 
accelerators~\cite{nowatzki2018hybrid,weng2020dsagen,weng2022unifying,liu2022overgen,weng2020hybrid,nowatzki2017stream} utilize instructions to schedule 
computation and memory accesses.

\vspace{-9pt}
\subsection{Acceleration Challenges \& Our Solutions}
\label{sec:accchallenge}

\subsubsection{How to support an arbitrary problem and terminate acceleration processing on the fly?}

Many previous FPGA
accelerators support a fixed-size
problem such as deep learning accelerators~\cite{wang2021autosa,cong2018polysa,zhang2015optimizing}, stencil computation~\cite{chi2018soda,chi2020exploiting}, graph convolutional network acceleration~\cite{zeng2020graphact}, 
and other applications~\cite{guo2019hardware}.
FPGA accelerators that support fixed-size
problems need to re-perform synthesis/place/route
flow for a new problem. The synthesis/place/route
flow takes hours to days to finish, which
is not suitable for the frequent invocation
for different problems in data centers.
Thus, the accelerator needs to support
an arbitrary problem once deployed. However,
the JPCG has six dimensions of freedom
as Algorithm~\ref{alg:cg} illustrates, which
makes 
it challenging to support
an arbitrary size problem. 

Another unique challenge is
designing an accelerator that is able to terminate 
on the fly. 
Because the JPCG will terminate
the main loop once the residual is less than
a preset threshold (Line 6 of Algorithm~\ref{alg:cg}), which we 
do not know until
we run the algorithm/accelerator.
In contrast,
in deep learning acceleration,
we know the iteration numbers of all loops
before the execution.

\noindent
\textbf{\calli Solution: Stream Centric Instruction Set.}
We design a stream centric instruction set
to control the processing modules and 
data flows in the accelerators.
We encode vector and matrix
size and data flow directions in the
instruction. 
A global controller is responsible
to issue instructions.
Therefore, we are able
to support an arbitrary problem for the JPCG
and
terminate the accelerator.
There are three principles for designing
our stream centric instruction set: 

\noindent\textbf{(1) Stream centric.} Every instruction is to process some streams.
The JPCG is dealing with vectors 
and matrices and we transfer those
in streams.
This principle naturally enables
task parallelism.

\noindent\textbf{(2) Data streamed processing.} We introduce a processing model that will either procure or consume streams in the accelerator. 
We use an instruction to control
the behavior of a processing module.

\noindent\textbf{(3) Decoupled memory and computing.} We separate the memory load/store
from computation. In this way, we can benefit from prefetching and overlapped computing and memory accessing.

Section~\ref{sec:arch} and Section~\ref{sec:instruction} will discuss details on the designs of 
\calli architecture, instructions, and processing modules.

\vspace{-6pt}
\subsubsection{How to coordinate long-vector data flow among processing modules?}

The vector length in a real-world
JPCG problem could be a few thousand
to more than
one million as shown in Table~\ref{table:matrices}.
Because we process floating-point values,
one vector size will be up to tens of megabytes
which exceeds the on-chip memory size. One straightforward 
way is that we always store an output
vector from a processing module to the off-chip 
memory and load an input vector from the off-chip 
memory to a module. However, there are reuse opportunities to save some off-chip memory load and store in JPCG. For example, the output vector
$\vec{\mathbf{ap}}$ by Line 7 will be consumed by
Line 8 and Line 10. But it is non-trivial to reuse
$\vec{\mathbf{ap}}$ for Line 10 during streaming 
because there is a dependence (i.e., $\alpha$) of Line 10 on Line 8,
and Line 8 will not produce $\alpha$ until
it consumes the whole vector $\vec{\mathbf{ap}}$.
As a result, we need to store the whole $\vec{\mathbf{ap}}$ on chip for reusing, which
is not practical. Therefore, we have a dilemma: 
(1) vector $\vec{\mathbf{ap}}$ 
exceeds the on-chip memory size so we need to store it in 
the off-chip memory, but
(2) we need to reuse $\vec{\mathbf{ap}}$ to reduce
off-chip memory accesses, however,
(3) the dependence issue requires us to store
the whole
vector $\vec{\mathbf{ap}}$ on chip or there will be no reuse.
Therefore, it is a challenge to coordinate
vector flows among processing modules for reusing
while resolving the dependence issue at the same time.

\noindent
\textbf{\calli Solution: Decentralized Vector Scheduling.}
In Section~\ref{sec:decentralized} we will introduce the concept of
vector streaming 
reuse and
analyze the dependency in the
JPCG and partition
the main loop in three phases so that
we will reuse 
vectors within the same phase via on-chip
streaming while store/load vectors to/from
memory across phases. Based on the
vector reusing/loading/storing, 
we will form a decentralized vector scheduling
to dissolve the global control to  
vector control and computation modules
to coordinate vector
flows among modules. 

\vspace{-12pt}
\subsubsection{How to save off-chip memory bandwidth and maintain double (FP64) precision?} 

To represent a double-precision (FP64) non-zero, we
need 32 bits for the row index, 
32 bits for the
column index, and 64 bits for the FP64 value.
So we need 128 bits to represent an FP64 non-zero.
Similarly, we need 96(=32+32+32) bits to represent an FP32 non-zero. 
The memory port has a limited bit width. For
example, the AXI bus width is up to 512 bits~\cite{cong2018best,choi2021hbm}.
Therefore, lower precision (less bits per data element) provides
higher parallelism.
However, the
default JPCG requires the
FP64 precision
for convergence. There are many vectors and 
one sparse matrix in the
JPCG. How can we configure the 
precision (FP32 or FP64) for the vectors and matrix
in the
JPCG to save memory bandwidth but also
converge as effective as
the default FP64
precision?
\begin{wraptable}{r}{0.25\textwidth}
  \vspace{-3pt}
  \caption{Three mixed-precision schemes for SpMV
  $\vec{\mathbf{y}} = \mathbf{A}\vec{\mathbf{x}}$.}
  \label{table:mixed}
  \vspace{-12pt}
  \footnotesize
  \centering
  \begin{tabular}{cccc}
    \hline
    & $\mathbf{A}$ & $\vec{\mathbf{x}}$ & $\vec{\mathbf{y}}$ \\
    \hline
    \textbf{Default FP64} & 
    FP64 & FP64 & FP64 \\
    
    \textbf{Mixed-V1} & 
    \color{blue}{FP32} & \color{blue}{FP32} & \color{blue}{FP32} \\

    \textbf{Mixed-V2} & 
    \color{blue}{FP32} & \color{blue}{FP32} & FP64 \\

    \textbf{Mixed-V3} & 
    \color{blue}{FP32} & FP64 & FP64 \\
    
    \hline
  \end{tabular}
  \vspace{-9pt}
\end{wraptable}

\noindent
\textbf{\calli Solution: Mixed-precision SpMV.}
Because the
JPCG refines vectors in the
main loop,
we must maintain all vectors in FP64 at the end of
each iteration. The SpMV takes as input
one vector $\vec{\mathbf{x}}$ 
and one sparse matrix
$\mathbf{A}$ and output a vector $\vec{\mathbf{y}}$.
Among the two vectors and one matrix,
the sparse matrix dominates the memory footprint. 
Thus, we consider 
using
FP32 for the sparse
matrix. For the SpMV input/output vectors, we have
two precision options -- FP32 or FP64.
Thus, we have three mix-precision schemes, illustated
in Table~\ref{table:mixed}.
Noted the mixed precision only applies to the SpMV,
and we always maintain the the vectors in the main
loop in FP64.
We will discuss the mixed precision in
\calli and the hardware design for supporting
mixed-precision SpMV in Section~\ref{sec:mixedspmv}.


\vspace{-6pt}
\section{\calli Architecture}
\label{sec:arch}

\begin{figure}[tb]
\centering
\includegraphics[width=0.55\columnwidth]{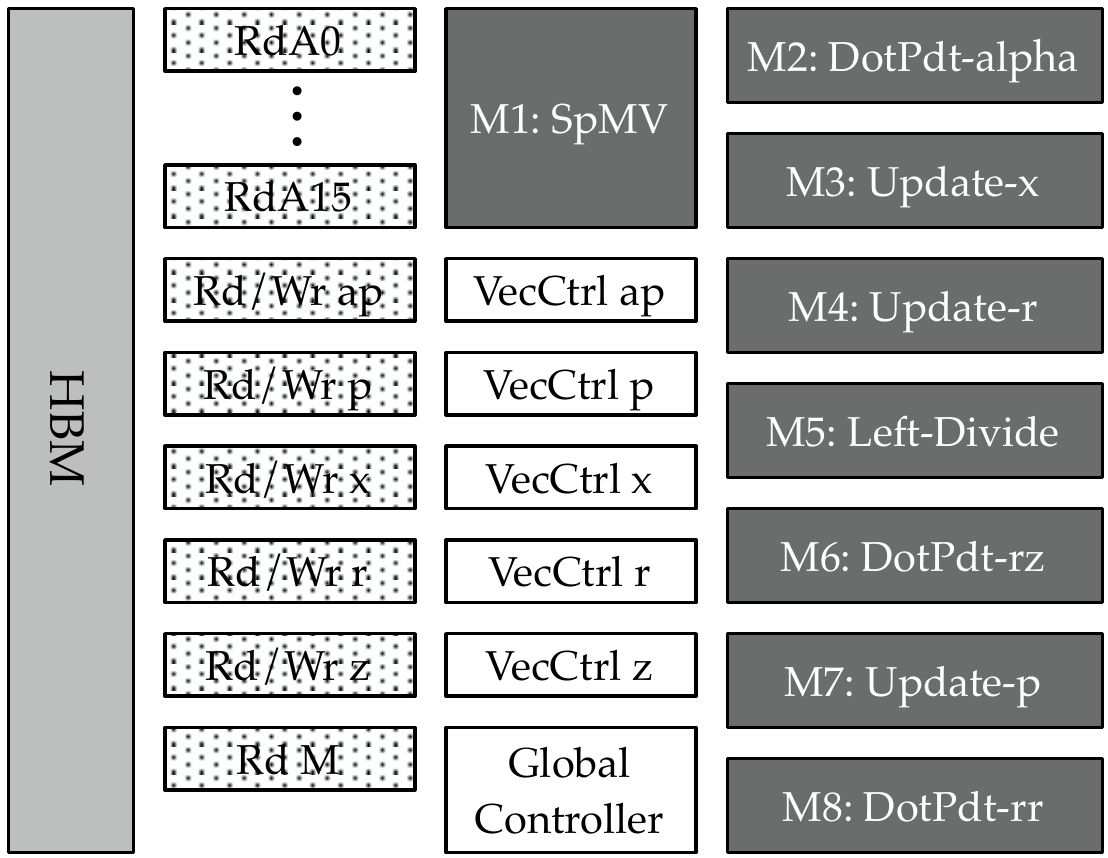}
\vspace{-12pt}
\caption{Top architecture of \calli accelerator.
}
\label{figure:arch}
\vspace{-9pt}
\end{figure}

\begin{figure}[tb]
\centering
\includegraphics[width=0.8\columnwidth]{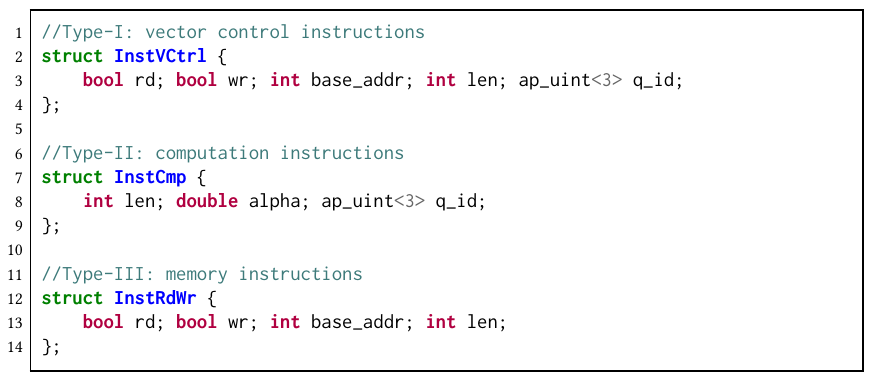}
\vspace{-12pt}
\caption{Three instruction types in \calli.
}
\label{figure:instruction}
\vspace{-15pt}
\end{figure}

Figure~\ref{figure:arch} shows the top
architecture of \calli accelerator
which is a modular architecture.
There are four categories of 
modules -- (1) computation
units, (2) read/write modules, (3) 
vector control modules, and (4) a global controller.
All modules are connected via FIFOs.

\textbf{Computation modules} perform
the vector/matrix computations. We have eight
computation modules -- 
(1) M1: SpMV, 
performing the computation of Line 7 in Algorithm~\ref{alg:cg},
(2) M2: dot product alpha, 
performing the computation of Line 8,
(3) M3: update x, performing the computation of Line 9,
(4) M4: update r,
performing the computation of Line 10,
(5) M5: left divide,
performing the computation of Line 11,
(6) M6: dot product rz,
performing the computation of Line 12,
(7) M7: update p, 
performing the computation of Line 13,
and
(8) M8: dot product rr
performing the computation of Line 15.
For Line 1 to Line 5, 
we reuse the eight computation modules
to perform the computation.
We leverage the open-sourced Serpens~\cite{song2022serpens}
accelerator for SpMV computation and
design the modules M2 to M8 for \calli.

\textbf{Memory read/write modules} move data from off-chip memory to
on-chip modules or vice versa.
We use the high bandwidth memory on Xilinx 
U280 FPGA as our off-chip memory.
We have sixteen read A modules (RdA0 to RdA15)
to read non-zeros to the SpMV module M1 and
a Rd M to read the Jacobi matrix.
There are five Rd/Wr (read-and-write) modules
for vectors $\vec{\mathbf{ap}}$,
$\vec{\mathbf{p}}$,
$\vec{\mathbf{x}}$,
$\vec{\mathbf{r}}$, and
$\vec{\mathbf{z}}$, because the five vectors
need both read and write operations.
We connect each read/write module
to one HBM
channel.

\textbf{Vector control modules} VecCtrl ap, p, x, r, and z coordinate
vector flows between one corresponding read/write modules to multiple computation units.
For example, according to Algorithm~\ref{alg:cg}, M1 (SpMV) produces
vector $\vec{\mathbf{ap}}$ and 
M2 (dot product alpha) and
M4 (update r) consume $\vec{\mathbf{ap}}$.
So the module VecCtrl ap coordinates
$\vec{\mathbf{ap}}$ vector flows among 
Rd/Wr ap and three
computation modules
M1, M2, and M4.

\textbf{The global controller} issues
instructions to computation modules
and vector control modules.
The global controller also performs some
scalar computation such as Line 14 of
Algorithm~\ref{alg:cg} and
compares $rr>\tau$ to decide
whether to terminate or not.

\vspace{-6pt}
\section{Stream Centric Instruction Set}
\label{sec:instruction}

\begin{figure}[tb]
\centering
\vspace{0pt}
\includegraphics[width=0.65\columnwidth]{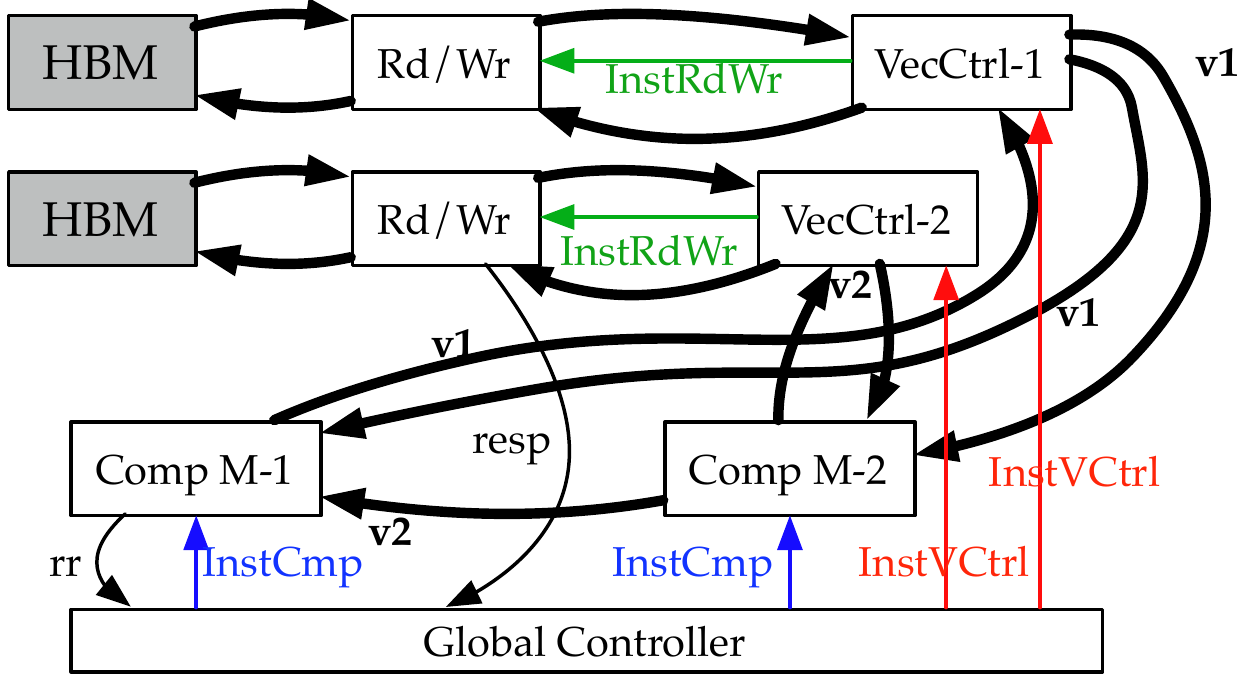}
\vspace{-12pt}
\caption{Processing model of stream based instructions.}
\label{figure:pmodel}
\vspace{-18pt}
\end{figure}

\subsection{Three Instruction Types}
We define the instruction types as 
illustrated in Figure~\ref{figure:instruction}
for \calli. They are 
(1) Type-I: vector control instructions,
(2) Type-II: computation instructions,
(3) Type-III: memory instructions.

\vspace{-6pt}
\subsubsection{Type-I: vector control instructions} We use
vector control instructions to tell
a vector control module where and how to
deliver a vector. Type-I instructions have 
five components -- (1) \texttt{int rd}
and (2) \texttt{int wr} encode
whether
to read or write or
simultaneously read and write
a vector, (3)
\texttt{int base\_addr} encodes the base address
of a vector in the memory, 
(4) \texttt{int len} encodes the vector length,
and 
(5) \texttt{ap\_uint<3> q\_id} encodes 
the index of a destination module
where to send the vector.

\vspace{-6pt}
\subsubsection{Type-II: computation instructions}
We use computation instructions to trigger 
the execution of a computation module
and where to send the output vector.
Type-II instructions have 
three components -- 
(1) \texttt{int len} encodes the vector length,
(2) \texttt{double alpha} a double-precision
constant scalar involved in the computation,
and
(3) \texttt{ap\_uint<3> q\_id} encodes 
the index of a destination module
where to send the vector. 
Note that the computation instructions
do not have operation code because
a computation module in the accelerator
only has one function.

\vspace{-6pt}
\subsubsection{Type-III: memory instructions}
We use memory instructions to read
a vector from off-chip memory to a vector control module or write a vector from 
a vector control module to 
off-chip memory.
Type-III instructions have 
three components -- 
(1) \texttt{int rd}
and (2) \texttt{int wr} encodes whether
to read or write or
simultaneously read and write
a vector,
(3) \texttt{int base\_addr} encodes the base address
of a vector in the memory, and
(4) \texttt{int len} encodes the vector length. 

\vspace{-6pt}
\subsection{Processing Model}
Figure~\ref{figure:pmodel} displays
an example where we process two 
vectors with two computation modules.
A global controller issues
vector control instructions
to the two vector control modules
VecCtrl-1 and VecCtrl-2. \textbf{v1}
and \textbf{v2} are two vectors, 
\texttt{rr} is a scalar, and
\texttt{resp} is a memory response.

\noindent
\textbf{Vector flow.} Instructions 
control vector flow
among modules in a streaming fashion.
For example,
if we are reading vector \textbf{v1}
with a length 100
from memory to Comp M-2, the 
controller will issue an
instruction \texttt{InstVCtrl\{rd=1, wr=0, base\_addr=0, len=100, q\_id=1\}}
to VecCtrl-1. Here, \texttt{q\_id=1} indicates
the destination module is M-2 rather than M-1.
Then VecCtrl-1 will
issue a memory instruction
\texttt{InstRdWr\{rd=1, wr=0, base\_addr=0, len=100\}}
to the memory module. Next,
vector \textbf{v1} flows from the
memory to Comp M-2. 
Another example is that the controller
issues a computation instruction
\texttt{InstCmp\{len=100, alpha=2.0, q\_id=1\}}
to Comp M-2. We assume M-2 performs
$\mathbf{v2} = \mathbf{v2} + \alpha\mathbf{v1}$.
Then M-2 will consume the input 
\textbf{v1} and \textbf{v2} vectors
and deliver the result 
vector \textbf{v2}(=\textbf{v2} + 2.0*\textbf{v1})
to M-1.

\noindent
\textbf{Scalar and memory response.} We update
all scalars in the global controller. For
instance, 
Comp M-1 delivers a scalar rr
to the controller and the controller
will decide whether to terminate the accelerator or not.
At the memory modules, 
we always send out a response to 
the controller if we are processing 
a memory write operation. The response
message will help the controller
to maintain memory consistency
when multiple modules read and write
the same vectors.

\noindent
\textbf{Overlapped execution and prefetching.}
The models involved are working
in parallel, i.e., task parallelism,
because we never cache a whole vector
on chip. One element in an input stream
will be consumed by a module and sent
to an output vector flow at each cycle.
So the modules in \calli accelerators
work with an II=1 pipeline\footnote{For the dot 
product modules, we have two phases. Phase I
multiplies input and accumulates
in a cyclic
delay buffer with an II=1 pipeline. Phase II
accumulates contents in the delay buffer with a
larger II=5 pipeline because of the read after write dependency and the FP ADD latency. The Phase II cycle count is fixed, i.e., $5*L$ where $L$ is the delay buffer size, for any arbitrary length vector. So the Phase II cycle count is 
negligible compared to the
Phase I cycle count.}.
We always prefetch vectors in the processing.
We enable prefetching by issuing
multiple
instructions.
For example, if Comp M-1 needs input vector
\textbf{v1}
from memory and input vector \textbf{v2}
from Comp M-2, along with the
computation instruction to M-2,
we also issue the vector control
instruction to the vector control
module to read \textbf{v1}
to M-1 for prefetching.

\noindent
\textbf{Processing rate matching.}
Prior work~\cite{cong2014combining}
processing rate matching of modules
in streaming applications. In this work,
because we do not cache any vector on
chip and we overlap the execution,
we match the vector input and output
streaming rate. Therefore, the bottleneck
becomes the connections between
a memory module and a memory channel. 
Although there are modules that use
multiple channels, we can simplify
the connection as one-module-one-channel
because we can view
a multi-channel module as multiple
one-channel modules and they are connected
via on-chip connections. Thus,
we can derive the accelerator frequency that
matches the memory bandwidth as
\begin{equation} 
\label{eq:freq}
f = BW / r,
\end{equation}
where $BW$ is per channel memory bandwidth
and $r$ is the maximum memory datawidth.
For a Xilinx U280 HBM FPGA~\cite{u280} which has
32 HBM channels and 460 GB/s memory bandwidth
and supports a
512-bit (64-byte) memory width~\cite{cong2018best,choi2021hbm},
the matching frequency is
$f = (460 GB/s / 32) / 64/B = 225$ MHz.

\vspace{-6pt}
\subsection{The Global Controller Code}
\begin{figure}[tb]
\centering
\includegraphics[width=0.7\columnwidth]{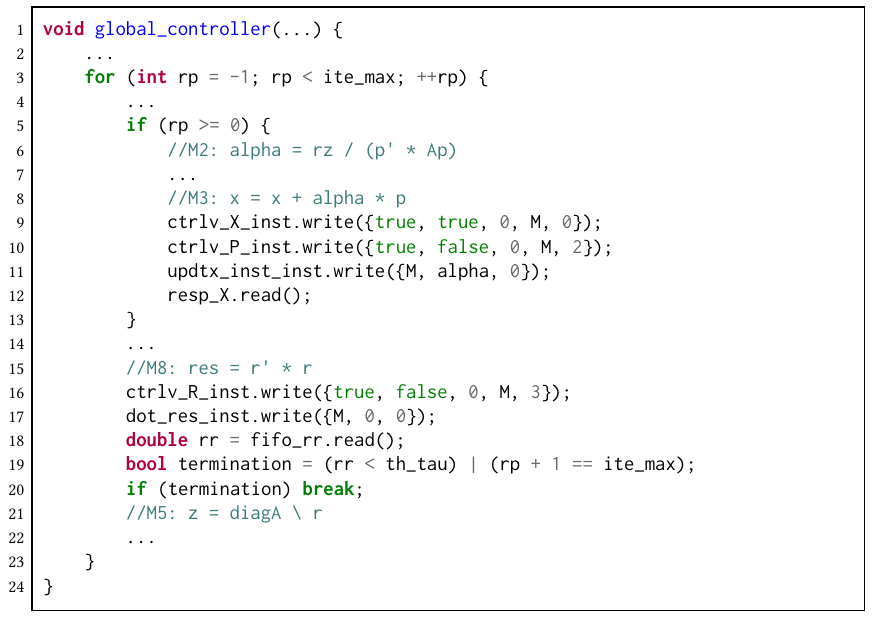}
\vspace{-12pt}
\caption{Controller code.}
\label{figure:controller}
\vspace{-15pt}
\end{figure}

Figure~\ref{figure:controller} shows
the instruction code in the global
controller. We only show
the main loop control code
and the intrusions for
vector control and computation for
module M3 and M8.
The code in Figure~\ref{figure:controller}
is quite similar to Algorithm~\ref{alg:cg}
because
\calli's instructions
make it easy for users
to control the
accelerators. In the controller code,
we have two optimizations -- (1) We
merge Line 1 to 5
of Algorithm~\ref{alg:cg} into the main
for loop to reuse the
modules. 
The if clause (Line 5 to 13)
in Figure~\ref{figure:controller}
skips some modules in iteration (rp=-1)
so that we can perform the computation
of  Line 1 to 5
of Algorithm~\ref{alg:cg} using the main 
for loop.
(2) We move the last module
M8 which is
the computation of residual 
before M5 to skip
the computations of M5 to M7 once
the solver converges.

\vspace{-6pt}
\section{Vector Streaming Reuse \& Decentralized Vector Scheduling}
\label{sec:decentralized}
\subsection{Vector Streaming 
Reuse}
The accelerator has to store long vectors to
off-chip memory because of limited 
on-chip memory size. However,
there are reusing opportunities
so that we can avoid unnecessary
load/store.
\textbf{Vector streaming 
reuse (VSR)} means vectors are reused
in a streaming fashion by processing modules
via on-chip streams/FIFOs. 

\noindent
$\bullet$\textbf{What is VSR?} A processing module (PM) consumes 
an element of a vector from an 
input stream/FIFO and produce 
an element (for a processed vector)
or duplicate an element (for the input vector) to an output stream/FIFO to another PM. The PMs
consume/produce vector elements
in pipeline and there may be multiple input/output streams/FIFOs
connected to one PM.

\noindent
$\bullet$\textbf{When can VSR?} (1) Multiple PMs consume the
same input vector(s), (2) a PM consumes vector(s) that are produced by some other PMs, and (3)
the difference of accessing indices of
two input vectors is within the on-chip (or stream/FIFO) memory budget.

\noindent
$\bullet$\textbf{When can not VSR?} (1) The computation of a scalar requires a whole vector
and PMs has dependency on the scalar can not reuse the vector and
(2) the difference of accessing indices of
two input vectors is out of the on-chip (or stream/FIFO) memory budget.

\vspace{-9pt}
\subsection{Three Computation Phases}

\begin{figure}[tb]
\centering
\includegraphics[width=0.95\columnwidth]{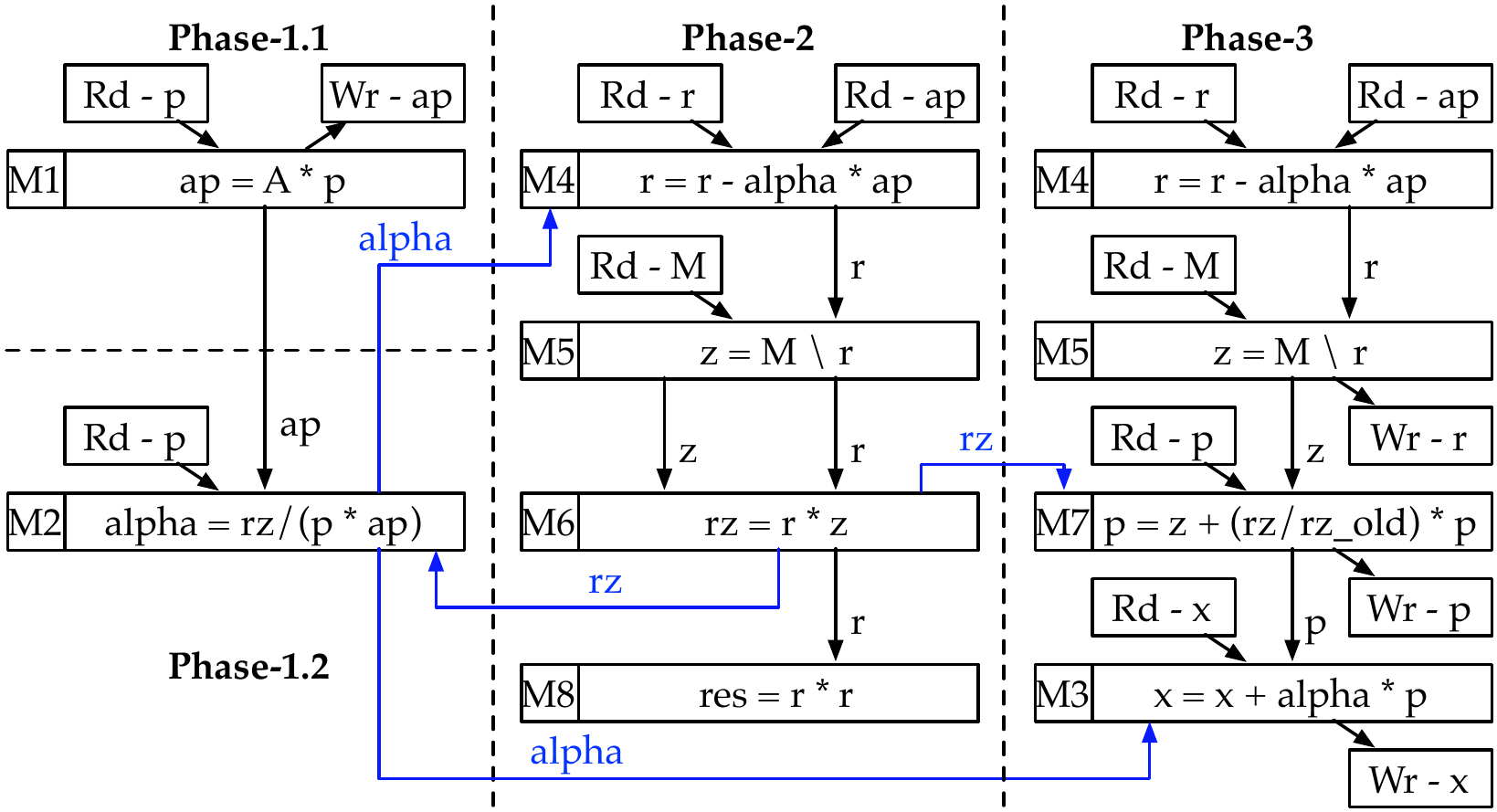}
\vspace{-12pt}
\caption{We divide computation
modules into three phases because of
the scalar dependency. Two key properties are -- 
(1) a scalar dependency separates
two phases
and (2) vectors within a phase can be reused
by modules in that phase.
}
\label{figure:phase}
\vspace{-15pt}
\end{figure}

We analyze the scalar dependency
and then divide the eight computation modules
into three phases as
shown in
Figure~\ref{figure:phase}.
The scalar dependency is the
critical issue that prevents us 
from
reusing vectors across computation
modules. 
For example, we can not reuse
the input vector $\vec{\mathbf{ap}}$ of M2
to M4 (which also takes $\vec{\mathbf{ap}}$
as input) because
M4 depends on alpha and alpha depends on
the whole vector $\vec{\mathbf{ap}}$.
So we move M4 to Phase-2.
Because M5/6/8 depends on vector $\vec{\mathbf{r}}$
from M4 so we direct
M5/6/8 to Phase-2. 
Similarly, we organize
the computations
in Phase-3 according to the dependency
on scalar rz.

\vspace{-9pt}
\subsection{Recomputing to Save Off-Chip Memory}

Among all vectors, vector
$\vec{\mathbf{z}}$ is a special one
because $\vec{\mathbf{z}}$ is an intermediate
vector in the main loop and it
is not reused between two iterations.
Thus, we reduce the off-chip
memory allocation for $\vec{\mathbf{z}}$
by recomputing in Phase-3. In
this way, we 
save memory channels.
Shown in Figure~\ref{figure:phase},
at Phase-2 after computation module
M5 produces $\vec{\mathbf{z}}$, we do not write
$\vec{\mathbf{z}}$ to memory. At Phase-3,
because M7 requires $\vec{\mathbf{z}}$
as input, so we reperform
both
M5 and M4.
M4 is the dependency of
M5 to produce vector  $\vec{\mathbf{z}}$
to M7.

\vspace{-9pt}
\subsection{VSR and Memory Accessing}

After forming the three computation phases,
we determine the VSR and 
the memory accessing for vectors
that we have to do. $\bullet$~\textbf{Phase~1:} In Phase-1.1
we perform M1 and in Phase-1.2 we perform
M2. We reuse the
$\vec{\mathbf{ap}}$ produced
by M1 to M7 to avoid reading 
the $\vec{\mathbf{ap}}$
from off-chip memory. We cannot reuse
vector $\vec{\mathbf{p}}$ from M1 for
M2 because
M1 outputs $\vec{\mathbf{ap}}$ only after consuming the whole vector $\vec{\mathbf{p}}$.
We write vector $\vec{\mathbf{ap}}$ to memory.
$\bullet$ \textbf{Phase 2:} We reuse vector $\vec{\mathbf{r}}$
by all the four modules M4/5/6/8. 
M4 consumes one entry from the input
stream of 
vector $\vec{\mathbf{r}}$ from memory and
immediately M4 sends the $\vec{\mathbf{r}}$
entry to the
next module M5. M5 and M6
perform the same consume-and-send
on vector $\vec{\mathbf{r}}$ so that
we only need to read vector
$\vec{\mathbf{r}}$ from memory once.
In this phase, we also have to
read vector $\mathbf{M}$ and vector $\vec{\mathbf{ap}}$ from memory once.
$\bullet$ \textbf{Phase 3:}
Similar to Phase 2,
M4 and M5 
reuse vector $\vec{\mathbf{r}}$
and M7
and M3 reuse
vector $\vec{\mathbf{p}}$. We have to read
vector $\vec{\mathbf{r}}$, $\mathbf{M}$,
$\vec{\mathbf{p}}$, and $\vec{\mathbf{x}}$ from
memory and write 
vector $\vec{\mathbf{r}}$, 
$\vec{\mathbf{p}}$, and $\vec{\mathbf{x}}$
to memory.


\vspace{-9pt}
\subsection{Decentralized Vector Scheduling}
\begin{figure}[tb]
\centering
\includegraphics[width=0.9\columnwidth]{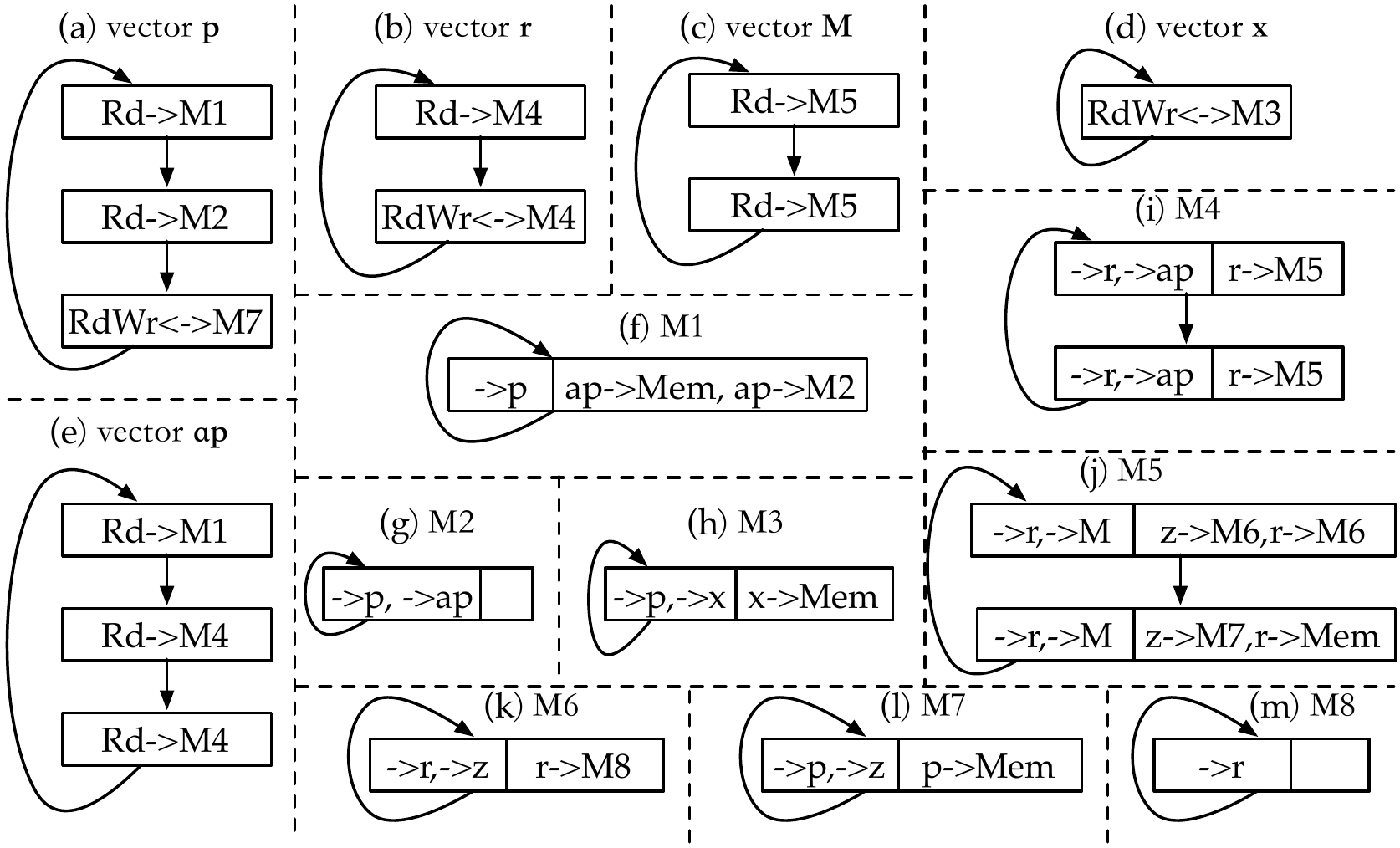}
\vspace{-12pt}
\caption{FSMs for decentralized vector scheduling in vector control modules (a) -- (e)
and computation modules (f) -- (m).
For vector control modules, Rd->Mx means to
read a vector from memory to computation module Mx, and RdWr<->Mx means 
to
read and write a vector to/from Mx from/to memory.
For computation modules, the left half block records input vectors and the right half block records output vectors and destination modules for VSR.}
\label{figure:vcschedule}
\vspace{-18pt}
\end{figure}
The VSR makes the control
complicated because we must
handle
both on-chip and off-chip vector
flows among all computation and
vector control modules.
We present decentralized vector scheduling
to relieve the pressure faced by
a centralized controller. Another
benefit is that decentralized vector 
scheduling is better for the controller 
routing because there are
23 FIFOs
for a centralized controller.
We dencentralize all vector scheduling
into each
individual vector control module 
(vector $\vec{\mathbf{p}}$, $\vec{\mathbf{r}}$,
$\mathbf{M}$, $\vec{\mathbf{x}}$, and $\vec{\mathbf{ap}}$)
and into the
computation modules 
(M1 to M8) show in 
Figure~\ref{figure:vcschedule}. 
We use a finite state machine (FSM)
to control the vector flow at each module.
Note that decentralized vector scheduling
maintains all dependencies.

\noindent
\textbf{Vector scheduling in vector control modules.} We use 
the FSM for $\vec{\mathbf{p}}$ in
Figure~\ref{figure:vcschedule} (a)
to illustrate
vector scheduling. 
According to Figure~\ref{figure:phase},
there are three memory operations
for  $\vec{\mathbf{p}}$: (1) Rd to M1 at Phase-1.1, 
(2) Rd to M2 at Phase-1.2, and
(3) Rd and Wr to/from M7 at Phase-3.
Thus, we have the FSM in Figure~\ref{figure:vcschedule} (a)
for $\vec{\mathbf{p}}$
scheduling.

\noindent
\textbf{Vector scheduling in computation modules.}
We use 
Figure~\ref{figure:vcschedule} (j)
the FSM for M5 to illustrate
the scheduling. 
According to Figure~\ref{figure:phase},
at Phase-2, M5 takes as inputs
the flows of the
vectors  $\mathbf{M}$
and  $\vec{\mathbf{r}}$, and outputs
$\vec{\mathbf{z}}$ and $\vec{\mathbf{r}}$ to M6,
resulting in
the first scheduling state.
At Phase-2, M5 uses
the flows of vector  $\mathbf{M}$
and  $\vec{\mathbf{r}}$ as inputs, 
but outputs
$\vec{\mathbf{z}}$ to M7 and $\vec{\mathbf{r}}$ to memory,
completing
the second scheduling state.

Without the decentralized vector scheduling, 
the accelerator accesses vectors 19 times (14 reads and 5 writes). With the decentralized vector scheduling,
the accelerator accesses vectors 14 times (10 reads and 4 writes).

\vspace{-9pt}
\subsection{Avoiding Deadlock}

\begin{figure}[tb]
\centering
\includegraphics[width=0.75\columnwidth]{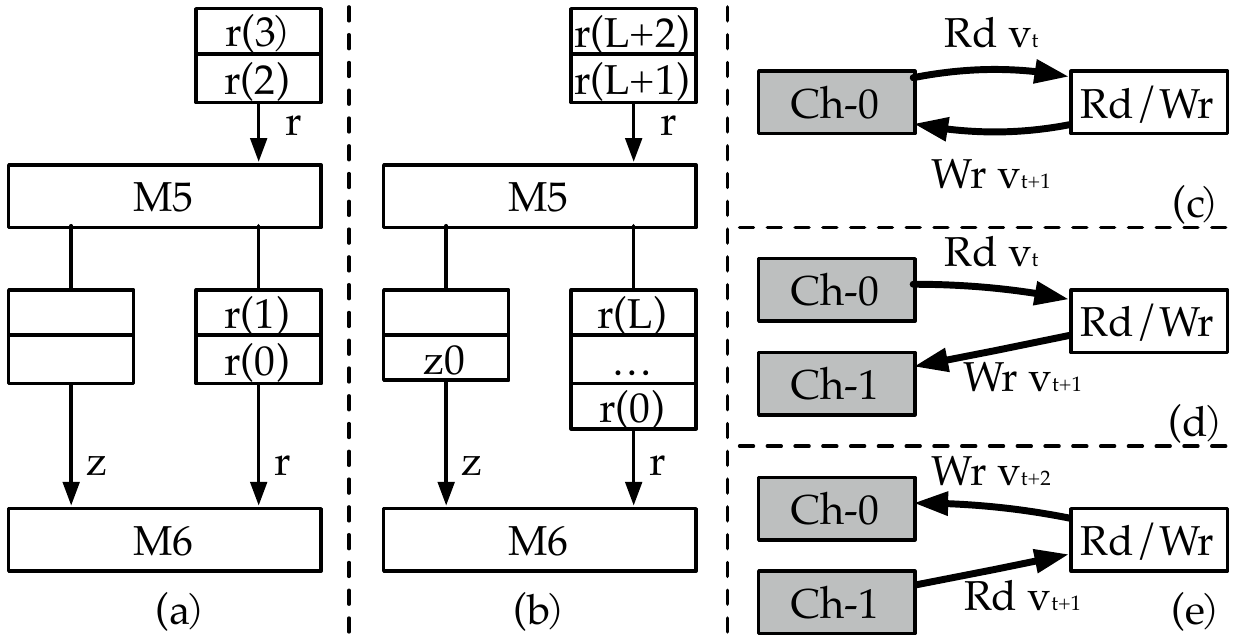}
\vspace{-12pt}
\caption{(a) A deadlock because of data arriving too late in slow FIFO, (b) increasing fast FIFO depth to resolve deadlock, (c) default one channel, (d) and (e) double channel design.}
\label{figure:deadlock}
\vspace{-18pt}
\end{figure}

A deadlock may occur when we
reuse more than one vector
to a destination module.
In Figure~\ref{figure:deadlock} (a),
the default FIFO depth is 2 and the 
M5 (left divide) pipeline depth is 
$L=33$.
Thus, when
the fast side $\vec{\mathbf{r}}$ FIFO is full
but the slow $\vec{\mathbf{z}}$ FIFO is still empty,
a deadlock occurs
because M5 cannot write
to $\vec{\mathbf{r}}$ FIFO and M6 cannot consume
$\vec{\mathbf{r}}$ FIFO entries (because
$\vec{\mathbf{z}}$ FIFO is empty).
To resolve the deadlock, we increase 
the fast FIFO depth to $>=L+1$ 
as shown in Figure~\ref{figure:deadlock} (b).
As a result,
during cycle 0 to $L$ M5
can write FIFO $\vec{\mathbf{r}}$ despite
FIFO $\vec{\mathbf{z}}$ is empty, but after
cycle $L+1$, M5 can write both FIFO
$\vec{\mathbf{r}}$ and $\vec{\mathbf{z}}$ and
M6 can read both FIFOs.

\vspace{-3pt}
\subsection{Double Channel Design}
By default a memory module reads and writes to 
the same memory channel
as shown in Figure~\ref{figure:deadlock} (c), which doubles the memory latency when
we perform both read and write on
a vector. Inspired by the widely
used on-chip double buffer design
in FPGA accelerators~\cite{zhang2015optimizing,wang2021autosa,sohrabizadeh2020end,zhao2017accelerating},
we present a double off-chip channel design.
In Figure~\ref{figure:deadlock} (d)
and (e),
we connect two channels to a memory module
and at the iteration $t$ we read vector 
$\vec{\mathbf{v}}_t$
from
channel 0 and write the updated 
$\vec{\mathbf{v}}_{t+1}$
to channel 1.
At the iteration $t+1$ we read
$\vec{\mathbf{v}}_{t+1}$
from channel 1
and write the updated 
$\vec{\mathbf{v}}_{t+2}$
to channel 0. Therefore, we reduce the
memory latency by half and maintain
the inter-loop vector dependency.


\vspace{-6pt}
\section{Mixed-Precision SpMV}
\label{sec:mixedspmv}

Table~\ref{table:mixed} presents three
mixed precision schemes and the default
FP64 precision. Overall, a scheme
with more data in FP64 is less
hardware efficient because it requires larger memory
capacity and higher memory bandwidth
but the accuracy is higher.
Mixed-V1 uses FP32 for all values
in the matrix and vectors. Although
Mixed-V1 is the most memory saving
scheme, it is
also the most inaccurate
scheme. 
Because the
JPCG is sensitive to
vector precision,
Mixed-V2 utilizes FP64 for the SpMV
output vector.
Mixed-V3 utilizes FP64 for both SpMV input and
output vectors.Among the three mixed-precision schemes,
Mixed-V3 does not sacrifice vector 
precision and at the same time saves
memory for the sparse matrix. 
Note that in SpMV the sparse matrix dominates the memory footprint.
Therefore, we use Mixed-V3
for the \calli accelerator for
both memory efficiency and computation accuracy.

\begin{wrapfigure}{r}{0.5\columnwidth}
\centering
\vspace{-6pt}
\includegraphics[width=\linewidth]{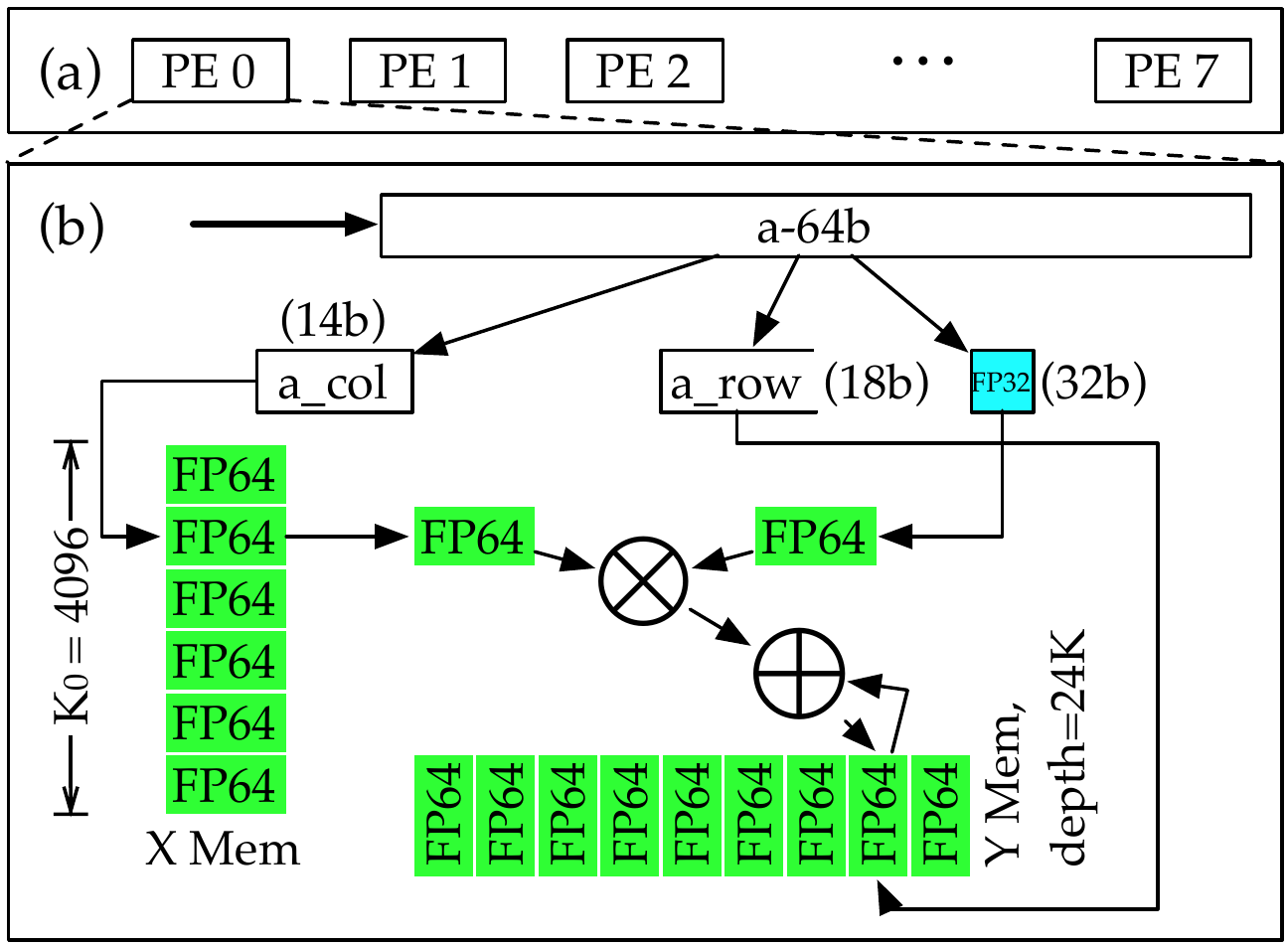}
\vspace{-12pt}
\caption{Mixed FP32/FP64 SpMV based on Serpens~\cite{song2022serpens}. 
}
\label{figure:pearch}
\vspace{-9pt}
\end{wrapfigure}
Figure~\ref{figure:pearch} illustrates 
the mixed-precision SpMV module architecture in \calli.
We leverage the Serpens~\cite{song2022serpens} architecture.
Each SpMV module is connected to one
memory channel and has eight
parallel processing engines.
The input to a processing engine
is a 64-bit element which contains
a 14-bit column index, an 18-bit row index,
and an FP32 value. For the mixed-precision
SpMV, we (1) store the input FP64 vector 
in an on-chip X memory implemented 
by BRAMs, and (2) buffer the output 
FP64 vector in an on-chip Y memory
implemented by URAMs. The depths
of the X and Y memories are 
4K and 24K respectively.
In the processing pipeline, we 
(1) cast the FP32 sparse value
into a FP64 value, (2) use
the column index to fetch the
corresponding
input element from the X memory,
(3) then
multiply the two FP64 scalars,
and (4) accumulate the result
to the
Y memory entry
indexed by row.

\vspace{-6pt}
\section{Evaluation}
\label{sec:evaluation}
\begin{table}[tb]
  \caption{The process node, frequency, HBM memory capacity, memory bandwidth, and power consumption of the NVIDIA A100, XcgSolver, SerpensCG, and \calli.}
  \label{table:platform}
  \vspace{-9pt}
  \footnotesize
  \centering
  \begin{tabular}{cccccc}
    \hline
    & Process & Frequency & Memory & Bandwidth & Power \\
    \hline
    
    \textbf{XcgSolver} & 
    16 nm & 250 MHz & 8 GB & 331 GB/s & 49 W \\
    
    \textbf{SerpensCG}  & 
    16 nm & 238 MHz & 8 GB & 345 GB/s & 43 W \\
    
    \textbf{\calli} & 
    16 nm & 221 MHz & 8 GB & 374 GB/s & 56 W \\

    \textbf{NVIDIA A100} & 
    7 nm & 1.41 GHz & 40 GB & 1.56 TB/s & 243 W \\
    \hline
  \end{tabular}
  \vspace{-9pt}
\end{table}

\begin{table*}[tb]
\caption{Matrix name, row/column number, and number of non-zeros (NNZ) of the evaluated matrices.}
\vspace{-9pt}
  \label{table:matrices}
  \centering
  \footnotesize
  \begin{tabular}{lrrr|lrrr|lrrr}
    \hline
    \textbf{ID} & \textbf{Matrix} & \textbf{\#Row} & \textbf{NNZ} &
    \textbf{ID} & \textbf{Matrix} & \textbf{\#Row} & \textbf{NNZ} &
    \textbf{ID} & \textbf{Matrix} & \textbf{\#Row} & \textbf{NNZ}\\
    \hline
M1 & {\tt ex9}	& 3,363	& 99,471 &
M2 & {\tt bcsstk15}	& 3,948	& 117,816 &
M3 & {\tt bodyy4}	& 17,546	& 121,550 \\
M4 & {\tt ted\_B}	& 10,605	& 144,579 &
M5 & {\tt ted\_B\_unscaled}	& 10,605	& 144,579 &
M6 & {\tt bcsstk24}	& 3,562	& 159,910 \\
M7 & {\tt nasa2910}	& 2,910	& 174,296 &
M8 & {\tt s3rmt3m3}	& 5,357	& 207,123 &
M9 & {\tt bcsstk28}	& 4,410	& 219,024 \\
M10 & {\tt s2rmq4m1}	& 5,489	& 263,351 &
M11 & {\tt cbuckle}	& 13,681	& 676,515 &
M12 & {\tt olafu}	& 16,146	& 1,015,156 \\
M13 & {\tt gyro\_k}	& 17,361	& 1,021,159 &
M14 & {\tt bcsstk36}	& 23,052	& 1,143,140 &
M15 & {\tt msc10848}	& 10,848	& 1,229,776 \\
M16 & {\tt raefsky4}	& 19,779	& 1,316,789 &
M17 & {\tt nd3k}	& 9,000	& 3,279,690 &
M18 & {\tt nd6k}	& 18,000	& 6,897,316
\\ \hline
\vspace{-6pt}
\\
\hline
M19 & {\tt 2cubes\_sphere} &	101,492	&	1,647,264 &
M20 & {\tt cfd2} &	123,440	&	3,085,406 &
M21 & {\tt Dubcova3} &	146,689	&	3,636,643 \\
M22 & {\tt ship\_003} &	121,728	&	3,777,036 &
M23 & {\tt offshore} &	259,789	&	4,242,673 &
M24 & {\tt shipsec5} &	179,860	&	4,598,604 \\
M25 & {\tt ecology2} &	999,999	&	4,995,991 &
M26 & {\tt tmt\_sym} &	726,713	&	5,080,961 &
M27 & {\tt boneS01} &	127,224	&	5,516,602 \\
M28 & {\tt hood} &	220,542	&	9,895,422 &
M29 & {\tt bmwcra\_1} &	148,770	&	10,641,602 &
M30 & {\tt af\_shell3} &	504,855	&	17,562,051 \\
M31 & {\tt Fault\_639} &	638,802	&	27,245,944 &
M32 & {\tt Emilia\_923} &	923,136	&	40,373,538 &
M33 & {\tt Geo\_1438} &	1,437,960	&	60,236,322 \\
M34 & {\tt Serena} &	1,391,349	&	64,131,971 &
M35 & {\tt audikw\_1} &	943,695	&	77,651,847 &
M36 & {\tt Flan\_1565} &	1,564,794	&	114,165,372 
\\ \hline
  \end{tabular}
  \vspace{-6pt}
\end{table*}
\begin{table*}[tb]

\caption{Solver time (in seconds) of the four accelerators: XcgSolver, SerpensCG, \calli, and A100 GPU. The speedup is the solver time of an accelerator/platform normalized to the XcgSolver solver time. We highlight an evaluation datum when it is the \textbf{\color{blue}{fastest among all four accelerators}} in \textbf{\color{blue}{blue}}
and an evaluation datum when it is \textbf{\color{red}{slower than the baseline
XcgSolver}} in \textbf{\color{red}{red}}. We also highlight the matrices where the
XcgSolver failed.
}
\vspace{-9pt}
  \label{table:solvertime}
  \centering
  \footnotesize
  \begin{tabular}{rllllllllllllllllll}
\cline{1-10}
& M1
& M2
& M3
& M4
& M5
& M6
& M7
& M8
& M9
\\ 
\cline{1-10}
\textbf{XcgSolver}(s) &
8.973E-1 &
4.151E-2 &
3.634E-2 &
3.825E-3 &
3.792E-3 &
5.219E-1 &
9.691E-2 &
1.268 &
3.577E-1 

\\ 

\cline{1-10}
\textbf{SerpensCG}(s) &
8.010E-1 &
2.787E-2 &
2.357E-2 &
2.656E-3 &
2.656E-3 &
4.217E-1 &
7.386E-2 &
1.245 &
2.719E-1

\\
Speedup &
1.120$\times$ &
1.490$\times$ &
1.542$\times$ &
1.440$\times$ &
1.428$\times$ &
1.238$\times$ &
1.312$\times$ &
1.018$\times$ &
1.315$\times$

\\
\cline{1-10}
\textbf{\calli}(s) &
\textbf{\color{blue}{2.602E-1}} &
\textbf{\color{blue}{9.200E-3}} &
\textbf{\color{blue}{6.579E-3}} &
\textbf{\color{blue}{9.261E-4}} &
\textbf{\color{blue}{9.376E-4}} &
\textbf{\color{blue}{1.408E-1}} &
\textbf{\color{blue}{3.020E-2}} &
\textbf{\color{blue}{4.213E-1}} &
\textbf{\color{blue}{1.021E-1}}

\\
Speedup &
\textbf{\color{blue}{3.449$\times$}} &
\textbf{\color{blue}{4.512$\times$}} &
\textbf{\color{blue}{5.524$\times$}} &
\textbf{\color{blue}{4.131$\times$}} &
\textbf{\color{blue}{4.045$\times$}} &
\textbf{\color{blue}{3.705$\times$}} &
\textbf{\color{blue}{3.209$\times$}} &
\textbf{\color{blue}{3.009$\times$}} &
\textbf{\color{blue}{3.502$\times$}}

 \\
\cline{1-10}
\textbf{A100}(s) &
\textbf{\color{red}{1.752}} &
\textbf{\color{red}{5.430E-2}} &
1.510E-2 &
3.681E-3 &
2.455E-3 &
\textbf{\color{red}{8.292E-1}} &
\textbf{\color{red}{2.076E-1}} &
\textbf{\color{red}{1.348}} &
\textbf{\color{red}{5.183E-1}}

\\
Speedup &
\textbf{\color{red}{5.120E-1$\times$}} &
\textbf{\color{red}{7.645E-1$\times$}} &
2.406$\times$ &
1.039$\times$ &
1.545$\times$ &
\textbf{\color{red}{6.294E-1$\times$}} &
\textbf{\color{red}{4.667E-1$\times$}} &
\textbf{\color{red}{9.407E-1$\times$}} &
\textbf{\color{red}{6.901E-1$\times$}}

\\

\hline \hline
& M10
& M11
& M12
& M13
& M14
& M15
& M16
& M17
& M18 
& \textbf{GeoMean}
\\ \hline
\textbf{XcgSolver}(s) &
1.613E-1 &
2.309E-1 &
3.336 &
3.333 &
4.540 &
1.246 &
4.883 &
3.813 &
1.018E+1 

\\ 

\hline
\textbf{SerpensCG}(s) &
1.162E-1 &
2.019E-1 &
\textbf{\color{red}{4.103}} &
2.983 &
\textbf{\color{red}{5.333}} &
1.050 &
\textbf{\color{red}{5.076}} &
3.238 &
7.970 

\\
Speedup &
1.389$\times$ &
1.143$\times$ &
\textbf{\color{red}{8.130E-1$\times$}} &
1.117$\times$ &
\textbf{\color{red}{8.513E-1$\times$}} &
1.187$\times$ &
\textbf{\color{red}{9.621E-1$\times$}} &
1.178$\times$ &
1.277$\times$ &
\textbf{1.194$\times$}
\\
\hline
\textbf{\calli}(s) &
\textbf{\color{blue}{4.103E-2}} &
\textbf{\color{blue}{7.104E-2}} &
\textbf{\color{blue}{1.488}} &
\textbf{\color{blue}{1.243}} &
\textbf{\color{blue}{1.872}} &
\textbf{\color{blue}{4.577E-1}} &
\textbf{\color{blue}{1.853}} &
1.580 &
3.785 
\\
Speedup &
\textbf{\color{blue}{3.932$\times$}} &
\textbf{\color{blue}{3.249$\times$}} &
\textbf{\color{blue}{2.242$\times$}} &
\textbf{\color{blue}{2.681$\times$}} &
\textbf{\color{blue}{2.425$\times$}} &
\textbf{\color{blue}{2.723$\times$}} &
\textbf{\color{blue}{2.636$\times$}} &
2.413$\times$ &
2.689$\times$ &
\textbf{\color{blue}{3.241$\times$}}\\
\hline
\textbf{A100}(s) &
\textbf{\color{red}{1.639E-1}} &
1.227E-1 &
2.074 &
1.298 &
1.903 &
6.153E-1 &
2.052 &
\textbf{\color{blue}{1.284}} &
\textbf{\color{blue}{1.924}} 

\\
Speedup &
\textbf{\color{red}{9.844E-1$\times$}} &
1.882$\times$ &
1.609$\times$ &
2.568$\times$ &
2.386$\times$ &
2.025$\times$ &
2.379$\times$ &
\textbf{\color{blue}{2.970$\times$}} &
\textbf{\color{blue}{5.291$\times$}} &
\textbf{1.395$\times$}
\\

\hline
\vspace{-3pt}

\\
\cline{1-10}
& M19
& M20
& M21
& M22
& M23
& M24
& M25
& M26
& M27
 \\ 
 \cline{1-10}
\textbf{XcgSolver}(s) &
1.004E-1 &
1.225E+1 &
9.410E-1 &
1.025E+1 &
\textbf{\color{red}{FAIL}}  &
1.187E+1 &
5.534E+1 &
3.291E+1 &
3.836

\\
\cline{1-10}
\textbf{SerpensCG}(s) &
2.956E-2 &
9.657 &
3.333E-1 &
7.436 &
4.984 &
9.353 &
5.055E+1 &
2.799E+1 &
3.138

\\
Speedup &
3.396$\times$ &
1.268$\times$ &
2.823$\times$ &
1.378$\times$ &
--- &
1.269$\times$ &
1.095$\times$ &
1.176$\times$ &
1.223$\times$

\\
\cline{1-10}
\textbf{\calli}(s) &
9.033E-3 &
2.928 &
1.039E-1 &
2.394 &
1.463 &
2.923 &
1.334E+1 &
7.558 &
1.056

\\
Speedup &
1.111E+1$\times$ &
4.182$\times$ &
9.053$\times$ &
4.280$\times$ &
--- &
4.061$\times$ &
4.150$\times$ &
4.355$\times$ &
3.632$\times$

\\
\cline{1-10}
\textbf{A100}(s) &
\textbf{\color{blue}{5.880E-3}} &
\textbf{\color{blue}{1.175}} &
\textbf{\color{blue}{5.671E-2}} &
\textbf{\color{blue}{9.354E-1}} &
4.183E-1 &
\textbf{\color{blue}{9.227E-1}} &
\textbf{\color{blue}{1.577}} &
\textbf{\color{blue}{1.081}} &
\textbf{\color{blue}{4.502E-1}} 

\\
Speedup &
\textbf{\color{blue}{1.707E+1$\times$}} &
\textbf{\color{blue}{1.043E+1$\times$}} &
\textbf{\color{blue}{1.659E+1$\times$}} &
\textbf{\color{blue}{1.095E+1$\times$}} &
--- &
\textbf{\color{blue}{1.287E+1$\times$}} &
\textbf{\color{blue}{3.511E+1$\times$}} &
\textbf{\color{blue}{3.045E+1$\times$}} &
\textbf{\color{blue}{8.522$\times$}} 

\\
\hline \hline
& M28
& M29
& M30
& M31
& M32
& M33
& M34
& M35
& M36
& \textbf{GeoMean}
\\ 
\hline
\textbf{XcgSolver}(s) &
\textbf{\color{red}{FAIL}}  &
1.956E+1 &
1.925E+1 &
\textbf{\color{red}{FAIL}}  &
\textbf{\color{red}{FAIL}}  &
\textbf{\color{red}{FAIL}}  &
\textbf{\color{red}{FAIL}}  &
\textbf{\color{red}{FAIL}}  &
\textbf{\color{red}{FAIL}}  
\\ \hline


\textbf{SerpensCG}(s) &
1.578E+1 &
1.189E+1 &
1.968E+1 &
6.738E+1 &
1.314E+2 &
3.134E+1 &
2.025E+1 &
1.021E+2 &
2.462E+2 

\\
Speedup &
--- &
1.645$\times$ &
9.783E-1$\times$ &
--- &
--- &
--- &
--- &
--- &
--- &
\textbf{1.490$\times$}

\\
\hline
\textbf{\calli}(s) &
5.508 &
4.548 &
6.291 &
2.277E+1 &
4.380E+1 &
1.044E+1 &
7.013 &
3.976E+1 &
8.970E+1 

\\
Speedup &
--- &
4.300$\times$ &
3.060$\times$ &
--- &
--- &
--- &
--- &
--- &
--- &
\textbf{4.787$\times$}

\\
\hline
\textbf{A100}(s) &
1.400 &
\textbf{\color{blue}{1.266}} &
\textbf{\color{blue}{1.227}} &
4.040 &
7.548 &
1.710 &
1.153 &
7.043 &
1.673E+1 

\\
Speedup &
--- &
\textbf{\color{blue}{1.545E+1$\times$}} &
\textbf{\color{blue}{1.568E+1$\times$}} &
--- &
--- &
--- &
--- &
--- &
--- &
\textbf{\color{blue}{15.72$\times$}}
\\ \hline
  \end{tabular}
  \vspace{-12pt}
\end{table*}

\subsection{Evaluation Setup}

\subsubsection{Benchmark Matrices} 
We evaluate on 36
real-word sparse 
matrices from SuiteSparse~\cite{davis2011university}.
Table~\ref{table:matrices} lists the name, row/ column number, number of non-zeros (NNZ),
and the ID used in this paper of each matrix.
Matrix M1 to M18 are the 
benchmarking matrices 
that the Xilinx Vitis 
HPC~\cite{hpcvitis,hpcvitis_github} XcgSolver used. 
Their row/column numbers rage from
3,920 to 23,052 and NNZ is up to 6.90 M.
To comprehensively evaluate the accelerators,
we select 18 more large-scale matrices
from SuiteSparse. The row/column numbers of
Matrix M19 to M36 span from 123 K to 1.56 M
and the NNZ is up to 114 M. The 36 sparse
matrices cover a wide range of applications
including structural problems, thermal problems,
model reduction problems, electromagnetics
problems, 2D/3D problems, and other engineering
and modeling problems.

We evaluate the Jacobi Preconditioner
Conjugate Gradient (JPCG) Solver. We set
the reference vector $\vec{\mathbf{b}}$
to an all-one vector and the initial 
$\vec{\mathbf{x}}$ to an all-zero vector.
We set the stop criteria as the residual $|\vec{\mathbf{r}}|^2 < 10^{-12}$. 
We also set a 20K maximum iteration number
no matter if
the solver converges or not.

\vspace{-9pt}
\subsubsection{Accelerators/Platforms}

We evaluate three FPGA JPCG
accelerators -- XcgSolver, SerpensCG, and
\calli and a GPU JPCG. 
We prototype all
three FPGA JPCG
accelerators on a Xilinx Alveo U280
FPGA~\cite{u280}. The GPU used is an NVIDIA A100.
Table~\ref{table:platform}
shows the specifications of the four evaluated
accelerators/platforms.
A100 GPU used more
advanced process node than that of 
U280 FPGA.

\noindent
\textbf{FPGAs.}
The design details of the three FPGA accelerators are:

\noindent$\bullet$ XcgSolver: XcgSolver is a JPCG solver from Xilinx Vitis HPC~\cite{hpcvitis,hpcvitis_github}. XcgSolver utilizes Vitis BLAS and SPARSE implementations for the SpMV and vector processing. We obtain the 
 source code from the Xilinx Vitis Libraries git repo. XcgSolver uses 
 the
 FP64 precision for all floating-point values. We use XcgSolver as
 a baseline in the evaluation.

\noindent$\bullet$ SerpensCG: We employ the Serpens~\cite{song2022serpens} for SpMV processing and modify Serpens to support 
the FP64 processing. The precision of all floating-point processing in SerpensCG is FP64. Although Serpens~\cite{song2022serpens} is a powerful SpMV(FP32) accelerator, it does not support 
the JPCG. So we build
SerpensCG as a strong baseline 
 to study the performance gap of 
 a JPCG accelerator based on
 Serpens SpMV accelerator with 
 minimum effort
 between Xilinx XcgSolver and 
 a fully optimized JPCG accelerator, i.e., \calli.
 Therefore, SerpensCG only leverages the stream based instruction
 set (presented in Section~\ref{sec:instruction}) for the
 JPCG without mixed precision or the vector related optimizations. 

\noindent$\bullet$ \calli: \calli is a fully
 optimized JPCG accelerator plus
 mixed precision (presented in Section~\ref{sec:mixedspmv}) and the vector related optimizations (introduced
 in Section~\ref{sec:decentralized}.
 We use
 Mix-V3 mixed precision where only the SpMV non-zero values are in FP32 and all 
 other processing is in FP64 for \calli.

All three accelerators allocate 16 HBM channels for SpMV non-zero processing. We 
build SerpensCG and \calli with TAPA framework~\cite{chi2021extending,guo2022tapa} and
leverage AutoBridge~\cite{guo2021autobridge}
for frequency boosting~\cite{guo2020analysis}. We use Xilinx
Vitis 2021.2 for back-end FPGA
implementation
for all 
three accelerators.
We utilize TAPA runtime to measure the
FPGA accelerator execution latency
and Xilinx Board Utility \texttt{xbutil}
to report the power information.

\noindent
\textbf{NVIDIA A100 GPU.}
We build a
GPU JPCG with CUDA version 11 on an NVIDIA A100 GPU.
We use cuSPARSE routine \texttt{cusparseSpMV} to compute SpMV
and cuBLAS routines \texttt{cublasDaxpy}, \texttt{cublasDscal},
\texttt{cublasDdot}, and
\texttt{cublasDcopy} for vector processing.
We measure the GPU execution time
with \texttt{cudaEventElapsedTime}
and the GPU power with
NVIDIA System Management Interface \texttt{nvidia-smi}.

The NVIDIA A100 GPU is much more powerful
than the FPGA accelerators as the specifications in Table~\ref{table:platform} show.
All four accelerators/platforms
use HBM2 for memory, but the A100
GPU memory is $4\times$ in terms of
capacity and $>4\times$ in terms of
bandwidth compared with the three FPGA 
accelerators. Meanwhile, the A100
GPU frequency is $5\sim6\times$ of the FPGA accelerator frequency.
However, in the following section we
will show that the FPGA accelerators
are able to outperform the GPU
in many aspects of the JPCG.

\vspace{-9pt}
\subsection{Solver Performance}

We compare the solver time 
of the 36 evaluated matrices
on the four accelerators/platforms
in Table~\ref{table:solvertime}.
The solver time is the measured
kernel time that a kernel
reaches the convergence criteria or
the maximum iteration number.
We also report the speedup which 
is defined as (the execution time of 
an accelerator/platform) / (the execution time of XcgSolver).

\vspace{-3pt}
\subsubsection{Matrix M1 to M18, the 18 medium-scale sparse matrices used by Xilinx Vitis.}
Overall, SerpensCG, \calli, and A100 GPU achieve $1.194\times$, 
$3.241\times$,  and $1.395\times$
geomean speedup compared with XcgSolver.
\calli is up to $5.524\times$
faster compared with XcgSolver
and  outperforms XcgSolver
on all 18 matrices.
Among the 18 matrices, 
\calli is the fastest and outperforms
A100 GPU on 16 matrices (M1 to M16).
SerpensCG achieves $1.194\times$
speedup compared with XcgSolver,
which indicates that
one can leverage the Serpens~\cite{song2022serpens} 
to support the FP64 JPCG 
with minimum efforts and 
realize a better performance than
Xilinx's XcgSolver.
If we compare \calli with
SerpensCG, \calli is 
$2.71\times$ faster
than SerpensCG. The performance
gain illustrates that there
is still speedup potential
although SerpensCG is faster than
XcgSolver, and
the mixed precision
and the vector related optimizations
leads to an even higher performance.
Meanwhile, \calli is $2.32\times$
compared with the A100 GPU performance.

\vspace{-3pt}
\subsubsection{Matrix M19 to M36, 18 large-scale sparse matrices.}
Overall, SerpensCG, \calli, and A100 GPU achieve $1.490\times$, 
$4.787\times$,  and $15.72\times$
geomean speedup compared with XcgSolver.
For the 18 large-scale matrices, we notice that (1) XcgSolver failed
on eight matrices because memory allocation exceeds available memory space while the other three
accelerators/platforms support
all the 18 large-scale matrices, and (2) \calli
achieved a
higher speedup than the speedup on Matrix M1 to M18 
compared with XcgSolver, i.e., 
$4.787\times$ v.s. $3.241\times$.
The superior speedup indicates that
\calli has better scalability and  
supports larger problem size than
Xilinx's XcgSolver.
The highest speedup \calli
achieves is $11.11\times$.
However, A100 GPU performs better
on Matrix M19 to M36 for the following 
reason.
The SpMV in CG is memory bound. The arithmetic intensity of an FP64 SpMV is 0.125 FP/B (in comparison, the arithmetic intensity of an FP64 dense 128-128 matrix-matrix multiplication is 10.7 FP/B). Thus, CG has low data reuse and demands high memory bandwidth for high performance. GPUs are good at high throughput processing. Thus, for smaller problems, it is difficult to utilize all computing resources and especially off-chip memory bandwidth for CG. So GPUs such as A100 which has a extremely high memory bandwidth (1.56TB/s) perform better on larger problems.

\vspace{-6pt}
\subsection{Computational \& Energy Efficiency}

\begin{table}[tb]
  \caption{Throughput, fraction of peak (FoP), and energy efficiency of the four accelerators/platforms. }
  \label{table:eff}
  \vspace{-6pt}
  \footnotesize
  \centering
\begin{tabular}{rcccc}
    \hline
\multicolumn{5}{c} {\textbf{Throughput -- GFLOP/s}} \\ \hline
    & Peak & Min & Max & GeoMean \\
    \hline
    \textbf{A100} & 
    \textbf{\color{blue}{29,200}} &  2.693 & \textbf{\color{blue}{179.8}} & \textbf{\color{blue}{29.53 (\textbf{4.379$\times$})}}\\
    
    \textbf{XcgSolver} & 
    410	& 2.065 & 19.43 & 6.743 (\textbf{1.000$\times$}) \\
    
    \textbf{SerpensCG} & 
    410 & 2.734 & 20.76 & 7.848 (\textbf{1.164$\times$})\\
    
    \textbf{\calli} &
    410 & \textbf{\color{blue}{10.36}} & 43.71 & 22.69 (\textbf{3.366$\times$})\\
    \hline
    \vspace{-6pt}
  \end{tabular}

\begin{tabular}{cccc}
    \hline
\multicolumn{4}{c} {\textbf{Fraction of Peak}} \\ \hline
\textbf{A100} & \textbf{XcgSolver} & \textbf{SerpensCG} & \textbf{\calli} \\
    0.616\% &
    4.74\% &
    5.06\% &
    \textbf{\color{blue}{10.7\%}}
    \\
    \hline
    \vspace{-6pt}
\end{tabular}

\begin{tabular}{rccc}
    \hline
\multicolumn{4}{c} {\textbf{Energy Efficiency -- GFLOP/J}} \\ \hline
    & Min & Max & GeoMean \\
    \hline
    \textbf{A100} & 
    1.108E-2 & 7.398E-1 & \textbf{\color{red}{1.215E-1	(0.883$\times$)}}\\
    
    \textbf{XcgSolver} & 
    4.214E-2 &  3.966E-1 & 1.376E-1 (\textbf{1.000$\times$})\\
    
    \textbf{SerpensCG} & 
    6.358E-2 &  4.827E-1 & 1.825E-1 (\textbf{1.326$\times$})\\
    
    \textbf{\calli} &
    \textbf{\color{blue}{1.851E-1}} &  \textbf{\color{blue}{7.806E-1}} & \textbf{\color{blue}{4.052E-1 (\textbf{2.945$\times$})}}\\
    \hline
\end{tabular}
  
  \vspace{-12pt}
\end{table}

Table~\ref{table:eff} shows
the throughput, 
fraction of peak (FoP),
and energy efficiency
of the three FPGA accelerators
and the A100 GPU.
We define throughput as (\# floating-point operations) / (solver time),
energy efficiency as
(throughput) / (power),
and FoP
as (maximum throughput) / (peak throughput).
For the
A100 GPU, we sum up the FP64
throughput of both Cuda cores and tensor cores from the report~\cite{a100} as the peak throughput, i.e., 26,200 GFLOP/s.
To estimate the peak throughput
of the
Xilinx U280, we synthesize
an 8-way parallel FP64 axpy module
and use the reported DSP number 88 to
estimate the DSP FP64 efficiency as
5.5 DSP/FLOP. Then we use the U280
DSP number 9,024 from~\cite{u280}
and a 250 MHz target freqeuncy
to estimate the U280 peak FP64
throughput as $9024 / 5.5 * 0.250 = 410$ GFLOP/s.

\vspace{-6pt}
\subsubsection{Throughput}
\calli achieves
22.69 GFLOP/s which is $3.366\times$
compared to Xilinx's XcgSolver.
For the lower bound,
\calli achieves 10.36 GLOP/s, the highest among all 
four accelerators/platforms
and is $3.85\times$ compared 
to the
A100 GPU. 
For the maximum throughput, 
\calli achieves 43.71 GLOP/s, higher than XcgSolver
(19.43 GLOP/s) but 
lower than the A100 GPU
(179.8 GLOP/s).
Because of
the stream based instructions
and decentralized vector scheduling,
\calli is efficient in controlling
the
processing modules. However,
for GPU, the kernel launching
control signal is issued from
the
host CPU, which leads to the inefficiency
of the
GPU when processing small-size
problems. So \calli achieves a higher
minimum throughput than the A100 GPU.
The FoP of A100 GPU, XcgSolver,
SerpensCG, and \calli are
0.616\%, 4.74\%, 5.06\%, and
10.7\%, respectively.
In fact, the
HPCG Benchmark~\cite{dongarra2015hpcg} uses the conjugate gradient
solver to benchmark computer clusters' performance. 
In the June 2022 Results of HPCG Benchmark, the FoP
ranges from 0.2\% -- 5.6\%.
The 10.7\% FoP achieved by
\calli is significant.

\vspace{-6pt}
\subsubsection{Energy Efficiency}
The A100 GPU, 
XcgSolver,
SerpensCG,
and \calli
achieve respectively
1.215E-1 GFLOP/J,
1.376E-1 GFLOP/J,
1.825E-1 GFLOP/J, and
4.052E-1 GFLOP/J in geomean 
energy efficiency.
\calli is $2.945\times$
compared with Xilinx's 
XcgSolver
and $3.335\times$
compared with the
A100 GPU.
\calli also
achieves the highest
minimum and maximum
energy efficiency.

\vspace{-6pt}
\subsection{Resource Utilization}

\begin{table}[tb]
  \caption{FPGA resource utilization of XcgSolver, SerpensCG, and \calli, all on the Xilinx U280 FPGA.}
  \label{table:ultilization}
  \vspace{-9pt}
  \footnotesize
  \centering
    \begin{tabular}{cccc}
    \hline
    & LUT & FF & DSP  \\
    \hline

    \textbf{XcgSolver} &
    503K (38.6\%) & 
    878K (33.7\%) & 
    1196 (13.3\%) \\

    \textbf{SerpensCG} &
    399K (30.6\%) &
    445K (17.1\%) & 
    1236 (13.7\%) \\

    \textbf{\calli} &
    509K (38.9\%) &
    557K (21.4\%)  &
    1940 (21.5\%) 

    \\
    \hline
    & BRAM & URAM \\
    \cline{1-3}

    \textbf{XcgSolver} &
    595 (29.5\%) &
    128 (13.3\%) \\

    \textbf{SerpensCG} &
    460 (22.8\%) &
    384 (40.0\%) \\

    \textbf{\calli} &
    716 (35.5\%)  &
    384 (40.0\%)

    \\
    \cline{1-3}
  \end{tabular}
  
  \vspace{-12pt}
\end{table}

Table~\ref{table:ultilization}
compares the utilization
of the FPGA resources including
LUT, FF, DSP, BRAM, and URAM
of the three FPGA accelerators.
Compared with XcgSolver,
\calli consumes almost the same LUT ($\sim$39\%) and
less FF (21.4\% v.s. 33.7\%).
\calli consumes more DSPs
(1940 v.s. 1196) than XcgSolver,
which indicates that
\calli has a higher computation
capacity. \calli uses
more BRAMs (716 v.s. 595)
and URAMs (384 v.s. 128).
In the \calli accelerator,
the SpMV requires 512 BRAMs
and all URAMs, the other 206 BRAMs
are consumed by Xilinx's add-on modules.

\vspace{-6pt}
\subsection{Iteration Number \& Residual Trace}

\begin{table*}[tb]
\caption{Iteration numbers of the evaluated matrices and the difference compared to CPU.}
\vspace{-12pt}
  \label{table:iteration}
  \centering
  \scriptsize
  \begin{tabular}{rllllllllllllllllll}
    \hline
& M1
& M2
& M3
& M4
& M5
& M6
& M7
& M8
& M9
& M10
& M11
& M12
& M13
& M14
& M15
& M16
& M17
& M18 \\ \hline
\textbf{CPU} &
20K &
634 &
164 &
26 &
26 &
9,441 &
1,713 &
15,692 &
4,821 &
1,750 &
1,266 &
20K &
12,956 &
20K &
5,615 &
20K &
9,904 &
11,816 \\ 
\hline
\textbf{XcgSolver} &
20K &
770 &
256 &
39 &
39 &
10,902 &
2,018 &
20K &
5,959 &
2,376 &
1,855 &
20K &
15,044 &
20K &
5,880 &
20K &
11,128 &
13,262 \\
Diff. to CPU &
0 &
+136 &
+92 &
+13 &
+13 &
+1,461 &
+305 &
+4,308 &
+1,138 &
+626 &
+589 &
0 &
+2,088 &
0 &
+265 &
0 &
+1,224 &
+1,446 \\
\hline

\textbf{\calli} &
20K &
635 &
164 &
26 &
26 &
9,491 &
1,705 &
20K &
4,824 &
1,749 &
1,265 &
20K &
13,109 &
20K &
5,611 &
20K &
9,903 &
11,823 \\
Diff. to CPU &
0 &
+1 &
0 &
0 &
0 &
+50 &
-8 &
+4,308 &
+3 &
-1 &
-1 &
0 &
+153 &
0 &
-4 &
0 &
-1 &
+7\\
\hline
\textbf{A100} &
20K &
633 &
164 &
26 &
26 &
9,246 &
1,716 &
1,5703 &
4,823 &
1,750 &
1,261 &
20K &
12,420 &
20k &
5,607 &
20K &
9,909 &
11,811 
\\
Diff. to CPU &
0 &
-1 &
0 &
0 &
0 &
-195 &
+3 &
+11 &
+2 &
0 &
-5 &
0 &
-536 &
0 &
-8 &
0 &
+5 &
-5 
\\ \hline
\vspace{-3pt}

\\
\hline
& M19
& M20
& M21
& M22
& M23
& M24
& M25
& M26
& M27
& M28
& M29
& M30
& M31
& M32
& M33
& M34
& M35
& M36 \\ \hline
\textbf{CPU} &
33 &
8,419 &
242 &
6,151 &
2,224 &
5,507 &
6,584 &
4,903 &
2,287 &
6,424 &
5,902 &
3,906 &
9,879 &
13,263 &
2,054 &
1,299 &
7,638 &
12,160 
\\  \hline
\textbf{XcgSolver} &
47 &
11,333 &
348 &
7,708 &
FAIL &
6,676 &
8,294 &
6,782 &
2,739 &
FAIL &
9,477 &
4,583 &
FAIL &
FAIL &
FAIL &
FAIL &
FAIL &
FAIL \\
Diff. to CPU &
+14 &
+2,914 &
+106 &
+1,557 &
--- &
+1,169 &
+1,710 &
+1,879 &
+452 &
--- &
+3,575 &
+677 &
--- &
--- &
--- &
--- &
--- &
--- \\
\hline

\textbf{\calli} &
33 &
8,458 &
242 &
6,150 &
2,222 &
5,525 &
6,584 &
4,916 &
2,285 &
6,424 &
6,040 &
3,893 &
9,829 &
13,259 &
2,053 &
1,314 &
7,656 &
12,163 \\
Diff. to CPU &
0 &
+39 &
0 &
-1 &
-2 &
+18 &
0 &
+13 &
-2 &
+0 &
+138 &
-13 &
-50 &
-4 &
-1 &
+15 &
+18 &
+3 \\
\hline
\textbf{A100} &
33 &
8,403 &
242 &
6,154 &
2,222 &
5,517 &
6,584 &
4,902 &
2,284 &
6,409 &
5,900 &
3,893 &
9,894 &
13,249 &
2,052 &
1,306 &
7,642 &
12,161
\\
Diff. to CPU &
0 &
-16 &
0 &
+3 &
-2 &
+10 &
0 &
-1 &
-3 &
-15 &
-2 &
-13 &
+15 &
-14 &
-2 &
+7 &
+4 &
+1
\\ \hline
  \end{tabular}
  \vspace{-9pt}
\end{table*}

\begin{figure*}[tb]
\centering
\includegraphics[width=1.75\columnwidth]{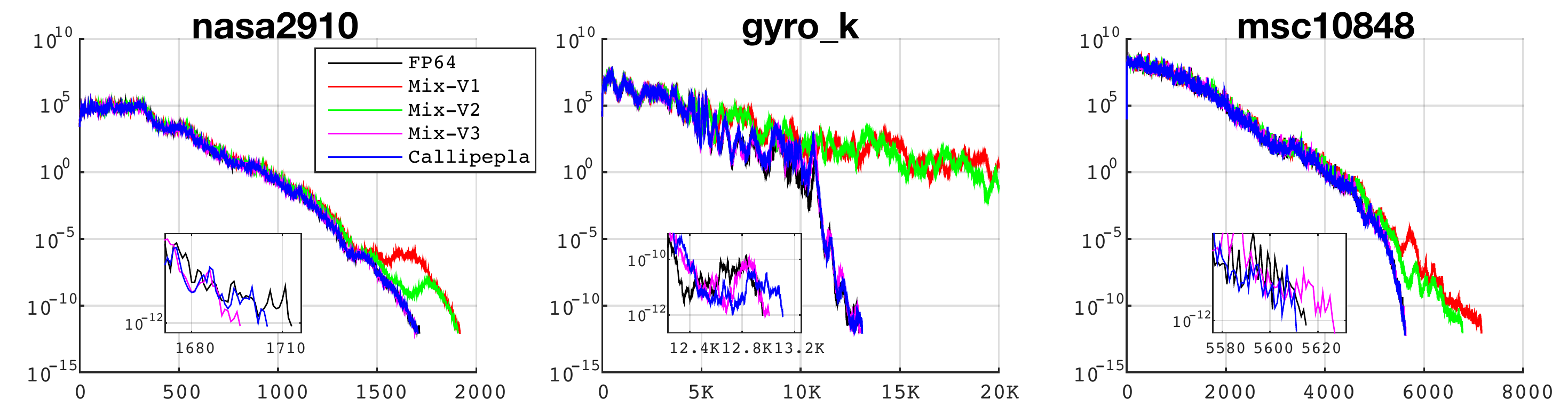}
\vspace{-12pt}
\caption{Residual traces of three matrices \texttt{nasa2910}, \texttt{gyro\_k},
and \texttt{msc10848} with five precision
settings:
default FP64, Mix-V1/V2/V3, and \calli on-board execution.
Y-axis: solver residuals; X-axis: iteration number.
}
\label{figure:trace}
\vspace{-15pt}
\end{figure*}

\subsubsection{Iteration Number} Table~\ref{table:iteration} reports
the iteration numbers of the evaluated 
matrices on the CPU,
Xilinx's XcgSolver, \calli, and
NVIDIA A100. 
We use the CPU  as
a golden reference.

For most matrices, the iteration
numbers of \calli and
the
A100 are within a
10 iteration difference compared to the
CPU.
However, XcgSolver shows 
significant iteration increases
on many matrices. For example,
on Matrix M20 \texttt{cfd2},
XcgSolver takes 2,914 more iterations to reach convergence.
XcgSolver pads zeros between
dependent elements in floating-point
accumulation to resolve the dependency issue. 
XcgSolver uses floating-point
accumulation latency as the dependency distance.
However, the HLS may insert extra latency when
scheduling the processing pipeline. 
Therefore, the true dependency distance
may become larger than the floating-point
accumulation latency. So
we observe the 
unstable numerical behaviors of XcgSolver.
On the contrary, the SpMV in \calli is based
on the Serpens~\cite{song2022serpens} accelerator
which uses the load-store dependency length instead
of the
floating-point accumulation latency. So the numerical
accuracy of \calli is higher than XcgSolver. 
Meanwhile, Serpens~\cite{song2022serpens} uses
an out-of-order scheme for scheduling non-zeros,
which can save memory and is reason that
XcgSolver exceeds available memory space and fails
on eight large-scale matrices while \calli
supports the eight matrices.

For Matrix M8 \texttt{s3rmt3m3}, both XcgSolver
and \calli do not converge but the CPU and the GPU
do. Because the two FPGA accelerators use
DSPs to implement FP64 multiplication and addition,
we suspect that there is a
numerical difference to some 
degree in the high-precision FP64 operation in 
the Xilinx 
HLS DSP implementation.

\vspace{-6pt}
\subsubsection{Residual Trace}

Figure~\ref{figure:trace}
illustrates
the residual traces of three matrices \texttt{nasa2910}, \texttt{gyro\_k},
and \texttt{msc10848} with 
default FP64, Mix-V1/V2/V3 precision
settings and the residual traces 
from \calli on-board execution.
From the residual trace of
\texttt{gyro\_k} we see that
Mix-V1 (where all values are in
FP32, colored in red) and Mix-V2 (where the SpMV input sparse matrix and input vector are in FP32, colored in green) do not converge
within 20K iterations. 
All the three residual traces
show the Mix-V3 (where only 
the SpMV input sparse matrix are in FP32, colored in magenta) are closely following the traces
of the default CPU FP64 (colored in black).
Although \calli employs the
Mix-V3 precision,
there is a small difference
of the trace from \calli on-board execution (colored in blue)
and the Mix-V3 trace. That is because
the difference in
hardware implementations of
the CPU FP64 processing
and the FPGA FP64 processing.

\vspace{-6pt}
\subsection{Bottleneck and Possible Improvement}
In the design of \calli accelerator,
we match the processing rate with
the HBM bandwidth as discussed in
Section~\ref{sec:instruction}. Therefore,
the bottleneck of \calli
is the HBM bandwidth. Exhibited in Table~\ref{table:platform} 
the bandwidth of the
NVIDIA A100 GPU
is $4.17\times$ of \calli (1.56 TB/s
v.s. 374 GB/s).
There are two HBM stacks on 
a Xilinx U280 FPGA for a total bandwidth
of 460 GB/s. 
If Xilinx deploys 8 ($4\times$) HBM stacks
on a next generation HBM FPGA,
we are able to achieve $3.07\times$
throughput advantage
compared to an A100 GPU.
However, the current HBM controllers
is area-hungry. In our evaluation,
the HBM controllers consume almost
one SLR and a U280 FPGA only has 3 SLRs. 
It is not practical to scale up $4\times$
bandwidth with the
current HBM controllers 
because that will consume 4 SLRs. 
We would like Xilinx to optimize the 
HBM controller IP or deploy it as an ASIC unit.

\vspace{-6pt}
\section{Conclusion}
In the design of FPGA JPCG accelerator we overcome three challenges -- (1) the support an arbitrary problem and accelerator
termination on the fly,
(2) the coordination of long-vector data flow among
processing modules, and
(3) saving off-chip memory bandwidth and maintaining FP64 precision convergence.
To resolve the challenges, we present \calli, an CG accelerator on Xilinx U280 HBM
FPGAs with
our innovative solutions -- (1)
a stream centric instruction set,
(2) vector streaming 
reuse and decentralized vector scheduling, and
(3) mixed FP32/FP64 precision SpMV.
The evaluation shows that compared to the Xilinx HPC product XcgSolver, \calli achieves a speedup of $3.94\times$, $3.36\times$ higher throughput, and $2.94\times$ better energy efficiency.
We also achieve 
77\% of the throughput with $3.34\times$ higher energy efficiency
compared
with an NVIDIA A100 GPU.

\vspace{6pt}
\noindent
\textbf{\large ACKNOWLEDGEMENTS}
\vspace{3pt}

\noindent
We thank the anonymous reviewers of FPGA 2023 for their constructive comments and Jinming Zhuang
for helping with the artifact evaluation.
This work is supported in part
by the NSF RTML Program (CCF-1937599),
CDSC industrial partners~(\url{https://cdsc.ucla.edu/partners}),
the Xilinx XACC Program,
and the AMD\footnote{J. Cong has a finacial interest in AMD.} HACC Program.


\bibliographystyle{ACM-Reference-Format}
\bibliography{references.bib}


\end{document}